\pdfoutput=1
\RequirePackage{ifpdf}
\ifpdf % We~are running pdfTeX in pdf mode
\documentclass[pdftex]{sigma}
\else
\documentclass{sigma}
\fi

\numberwithin{equation}{section}

\newtheorem{Theorem}{Theorem}[section]

 { \theoremstyle{definition}
\newtheorem{Definition}[Theorem]{Definition}

 }

\usepackage{tikz}
\usetikzlibrary{trees}
\tikzstyle{every picture}=[node distance = 10mm, baseline=-0.5ex]
\tikzstyle{every node}=[inner sep=2pt]
\tikzstyle{bla}=[shape=circle,minimum size=1.6mm,draw =black]
\tikzstyle{noi}=[shape=circle,minimum size=1.4mm, fill = black]
\tikzstyle{ver}=[shape=circle,minimum size=0.9mm, fill = black]
\tikzstyle{prop}=[shape=circle,minimum size=2.8mm, inner sep=0pt, draw=black!80, fill=green!30]
\tikzstyle{ ordvertex} = [shape=semicircle, draw=black, fill=brown!60, draw=brown!60!black]
\tikzstyle{twoint}=[shape=rectangle, minimum height = 16mm, fill=brown!40, draw = brown!40!black]

\def\dirp{{\slash \mkern -9mu p}}
\def\cI{{\mathcal I}}

\let\veps\varepsilon
\DeclareMathOperator{\id}{Id}
\DeclareMathOperator{\Tr}{Tr}

\begin{document}
\allowdisplaybreaks

\newcommand{\arXivNumber}{2207.05028}

\renewcommand{\PaperNumber}{007}

\FirstPageHeading

\ShortArticleName{Numerators in Parametric Representations of Feynman Diagrams}

\ArticleName{Numerators in Parametric Representations\\ of Feynman Diagrams}

\AuthorNameForHeading{M.P.~Bellon}

\Author{Marc P.~BELLON}
\Address{Sorbonne Universit\'e, CNRS, Laboratoire de Physique Th\'eorique et Hautes Energies,\\ Paris, France}
\Email{\href{mailto:marc.bellon@upmc.fr}{marc.bellon@upmc.fr}}

\ArticleDates{Received May 23, 2024, in final form January 24, 2025; Published online February 06, 2025}

\Abstract{The parametric representation has been used since a long time for the evaluation of Feynman diagrams. As a dimension independent intermediate representation, it allows a~clear description of singularities. Recently, it has become a~choice tool for the investigation of the type of transcendent numbers
appearing in the evaluation of Feynman diagrams. The inclusion of numerators has however stagnated since the ground work of Nakanishi. I here show how
to greatly simplify the formulas through the use of Dodgson identities. In the massless case in particular, reduction to the completion to a vacuum graph allows for a~strong reduction of the maximal power of the Symanzik polynomial in the denominator.}


\Classification{81T18; 81Q30; 15A15}

\section{Introduction}

In quantum field theories others than the scalar ones, propagators and vertices introduce numerators
depending on the various momenta, internal or external, in the integrals described by Feynman diagrams.
Many textbooks dismiss such numerators as inessential complications, but practical computations in
realistic theories soon encounter the formidable task it is to really compute those numerators. Even for
lowest order computations, with only tree diagrams, amplitudes can be tricky to evaluate.

Parametric representations have been an important tool for perturbative computations
in quantum field theory since their introduction by Feynman and Schwinger.
They are particularly interesting to prove general properties of the amplitudes and have been important
in many studies of renormalizability. More recently, they have been instrumental in the proof that a~whole
category of diagrams can be evaluated in terms of multiple zeta values (MZV)~\cite{Br2009,Brown_2008}.
A specialized program has been developed, which allows for the complete evaluation of the corresponding integrals~\cite{Pa13}.

The difficulty in having a simple representation of the numerators has hindered the
extension of such results to field theories with non-scalar fields. Nakanishi~\cite{Na71} worked out the
expression of the correlations between the momenta in two arbitrary edges in terms of graph polynomials, but the
straightforward application of these formulas results in far from optimal expressions, with really high powers of the first
Symanzik polynomial of the graph in the denominator. An important application of this formalism has been made in the work of Cvitanovi{\'c} and
Kinoshita~\cite{CvKi74}: they worked out parametric representations for the anomalous magnetic moment of the electron up to three loops,
even using quadratic Dodgson identities to simplify certain expressions.

However, most present evaluations of amplitudes avoid this kind of computations. Indeed, in gauge theories, a single topological graph can correspond
to multiple integrals, since, for example, a single three gluons vertex expands to three different terms. The usual route is therefore to reduce these
different terms to a basis of master integrals, which have no longer any numerators. Additional edges are added to the
diagram, but without introducing new loop momenta, such that all possible scalar products can be expressed as combinations
of the squared momenta in all edges of these extended diagrams. The diagram with its numerator can be expressed as a~combination of diagrams with positive or negative powers of the propagators corresponding to all these edges.
The whole family of such integrals is linked by integration by part relations (IBP methods), in order
to express all necessary diagrams with their numerators in terms of master integrals. A recent addition
has been to include higher dimension scalar integrals, allowing for a~set of integrals which are
completely finite and therefore easier to evaluate~\cite{MaPaSch14}. This program has been quite fruitful
and is at the basis of such computer codes as Mincer~\cite{mincer1,mincer2} or more recently
Forcer~\cite{UeRuVe16},
which allow for the evaluation of any three or four loop single scale massless diagram.\looseness=-1

Another direction has been the realization that some amplitudes in gauge theories are much simpler
than their individual components, starting from the maximally helicity violating (MHV) amplitudes. A
whole machinery has been set up which uses twistor representations of the momenta, recursion relations
to circumvent the evaluation of Feynman diagrams. The extension to loop amplitudes, even if it is
mainly limited to maximally supersymmetric theories, has similarly produced numerous important works.
It would be difficult to do justice to these numerous developments. A recent review of all these approaches as well as other ones can be found
in the papers of the SAGEX collaboration, which are presented in~\cite{sagex}.

The interest for the parametric representation remains for the studies of divergences and singularities. A recent example
is the study~\cite{GKNT23}, where the parametric representation is an essential tool for the disentanglement of the different types of divergence,
ultraviolet, infrared or associated to Landau singularities. Steps have also been taken to express the sum of numerators appearing in Yang--Mills
theory through the introduction of corolla polynomials~\cite{KrSaSu12}, but falling short to an explicit expression only in terms of
the Schwinger parameters.

Finally, a special place must be given to the work of Marcel Golz, who has made parallel progress on the explicit
parametric expression of a class of diagrams in QED, using some of the same identities~\cite{Go2019}. Using a
description of the Dirac trace problem with chord diagrams, he is able to evaluate this part of the problem for
a large number of propagators, even if it is in four dimension for a single fermionic loop. However, he does not look
for a reduction to the completed diagram, missing what is a very strong selling point of this work. The terms proportional
to external momenta remain in their
original form with a maximal power of the denominator $n+2$ for $n$ fermionic propagators instead of the
$n/2+2$ obtained in this work. This higher degree in the denominator would make the integration steps, either symbolic or numeric, much more
difficult.\looseness=-1

I will show that a parametric representation is possible which is nicer than the one given by the straightforward
application of Nakanishi's rules. The principal tool is a set of identities among graph polynomials stemming from
the Dodgson identities.\footnote{Some persons object to this name, since Dodgson only used the case of the dimension 2
determinant in his condensation method for the computation of determinants~\cite{Do1867}. Jacobi is at the origin of this
formula and is dutifully referred to by Dodgson. However, anything named after Jacobi requires further precisions, in
view of the range of his mathematics interests, so that I will stick to this name, used by a number of recent authors.
Furthermore, I cannot resist this occasion to refer to the author of {\em Alice in Wonderland}.}
I will use ideas and notations from Francis Brown's work~\cite{Br2009,BrYe09}. One interesting aspect of this rather direct approach
is that it avoids the ``black box'' aspect of IBP computations, which makes it difficult to have estimates on the integrals.
In application to the Wess--Zumino model, one obtains direct cancellation of divergences at the level of the integrand.
This is important for the application I envision, where Schwinger--Dyson equations are written directly in terms of renormalized amplitudes,
without any regulator.\looseness=-1

For scalar massless theories, the integrals only depend on a vacuum
graph obtained by closing the two point functions by an additional edge or joining all
exterior legs to an additional vertex, which is named the completed graph. I will first show how this dependence on only
the completed graph can be generalized to the case of a propagator graph with a single scalar product. The general case
would be difficult to solve using the same method, but a workaround is possible using scale invariance for propagator graphs.
In the case of fermion loops, expanding the traces in terms
of scalar products would produce far too many terms. It is better to start from
correlations of momenta and a nice reduction of the number and complexity of the terms
will be shown in the cases of loops of length four and six.

As a showcase for the power of this approach, I will deal with the
three and four loops primitive contributions to the two point functions in the supersymmetric
Wess--Zumino model, obtaining parametric expressions with the full symmetry of the
completed graph and no spurious divergences.

We limit ourselves here to the case of corrections to propagators for scalar particles, for which the reduction to vacuum diagrams is
straightforward. In the case of propagators for non-scalar particles, due
attention has to be taken to the different possible tensor structures while vertex type diagrams have their own
difficulties. In all cases, reduction to completed diagrams should be possible and allow
for similar simplifications in the parametric representations, at least in the massless case.

\section{Presentation of the main results} \label{presentation}

The notations for graph polynomials will be the ones in the works of
Brown and collabora\-tors~\cite{Br2009,BrYe09}. To each connected graph~$G$ a matrix $\tilde M_G$ is associated.
The matrix $\tilde M_G$ is a block
matrix with a diagonal block with the Schwinger parameters $x_i$ associated to the edges as diagonal values, two
blocks made of the incidence matrix and minus its transpose and a zero block to complete it. Schematically, it can be written
\begin{equation*}
	\tilde M_G = \begin{pmatrix}
		\operatorname{Diag}(\mathbf{x}) & -{}^t \operatorname{Inc} \\ \operatorname{Inc} & 0
	\end{pmatrix}.
\end{equation*}
This matrix is singular since the sum over the rows indexed by the vertices of the incidence matrix is zero, but the
determinant of any matrix $M_G$ obtained by removing the row and column corresponding to any vertex gives the first
Symanzik polynomial of the graph~$U_G$, independently of the chosen vertex. It is further independent of any of the choices made in the
presentation of the matrix, like the order of the edges and vertices or the orientation of the edges. This polynomial is of degree~$L$, with
$L$ the dimension of the first homology group of the graph, the loop number in the language of physicists.

The objects which will be used to express our results are the polynomials
$U^{I ,J}_K$ defined for~$I$,~$J$ and~$K$ subsets of the edges of the graph
such that $I$ and~$J$ have the same number of elements~$k$. The polynomial
$U^{I ,J}_K$ is the determinant of the matrix deduced from the graph matrix $M_G$ by
removing the rows associated to $I$, the columns associated to $J$ and setting
the variables associated to $K$ to zero and has degree $L-k$. From the symmetry properties of $M_G$,
one deduces that $U^{I ,J}_K = U^{J,I}_K$. These graph polynomials may change sign but are otherwise independent of
the choices made in the presentation of the graph. Since
we will need a definite sign for these expressions, it will be preferable to define
$U^{I ,J}$ as the determinant of the matrix where the rows indexed by $I$ are
replaced with rows with only one non zero value 1 in a position corresponding to a row in
$J$. Expansion along the columns indexed by~$J$ or the rows indexed by~$I$ gives back the previous definition up to
a sign, but now the result only depends on the orientation of the edges indexed by
$I$ and $J $ and the bijection between these subsets
defined by the positions of the 1 in the modified matrix. Schnetz in~\cite{SC2021} proposed to put
an explicit sign in front of the determinant of the reduced matrix. My implicit definition turns out to be fully equivalent to his
explicit one. Both approaches can be useful.

These polynomials, in the special case where $I$ and $J$ have only one element,
give the adjoint elements of the matrix $M_G$, so that the correlation between momenta in rows
$i$ and $j$ is given by $U^{\{i\},\{j\}}/U=U^{i,j}/U$ (we do not keep the braces in the case of
singletons to lighten the notations). The Dodgson identities relate determinants of matrices whose elements
are $U^{i,j}$ for~${i \in I}$ and $j \in J$ to the polynomials $U^{I ,J}$. In Appendix \ref{polyn}, I state these
identities and their proofs, paying due attention to the question of signs.

Our results is most easily described for a vacuum diagram. The case of a propagator diagram can be reduced to this one, as will be seen in next
section.
The graph is replaced by the related completed graph,
obtained by linking the two exterior legs in a simple edge. In this case, the exponent associated to the added edge
must be chosen to make the whole vacuum diagram scale invariant.

We consider a numerator expressed as the product for $s=(s_0,s_1)$ in a set~$S$ of $n$ scalar
products $p_{s_0} \cdot p_{s_1}$. A priori, there are no restrictions to the number of times the momentum
associated to a given edge can appear in this product of scalar products. It should be even possible to have higher
powers of some of the scalar products, but it seems to enter the domain of useless generality.
\begin{Definition}
For a non-empty subset~$S_0$ of~$S$, define $U^{S_0}$ as the mean of the determinants defined using
functions $\veps$ from~$S_0$ to the set of edges of the graph such that
$\veps(s) \in \{s_0,s_1\}$, together with the complementary function
$\tilde\veps$ such that $\{\veps(s),\tilde\veps(s)\} =
\{s_0,s_1\}$
\[
	U^{S_0} = \frac1 {2^{\#(S_0)}} \sum_{\veps} U^{\veps(S_0), \tilde\veps(S_0)}.
\]
\end{Definition}
Since the symmetry of the matrix means changing $\veps$
to $\tilde\veps$ does not change the term, $U^{S_0}$ has at most
$2^{\#(S_0)-1}$ different terms. With these objects, we can easily state our main theorem
\begin{Theorem}\label{fund}
The integrand of a massless vacuum diagram in space-time dimension~$d$ with numerator expressed
as a product of $n$ scalar products can be written as a sum over the set
$\mathcal P_k$ of partitions of~$S$ in $k$~parts, with $k$ taking all values for which this set is
non empty
\begin{gather*}
	I = \sum_{k=1}^{\#(S)} \frac{(-1)^{\#(S)-k}\Gamma\bigl(\frac d2 + k\bigr)}{U^{\frac d2 +k} }
			\sum_{\{S_1,\ldots,S_k\} \in \mathcal P_k} \prod_{j=1}^k U^{S_j}.
\end{gather*}
We can observe that this rational function has homogeneity $-L D/2 - n$.
The value of the residue of the diagram is then the projective integral obtained by multiplying $I$ with the
product~${\prod_i \frac 1{\Gamma(\beta_i)} x_i^{\beta_i-1} {\rm d }x_i^{\vphantom{\beta_i}}}$,
indexed on all edges of $\tilde G$. Since it is a projective integral, any delta function fixing a sum
of~$x_i$ give the same result. The exponents $\beta_i$ associated to the edges must sum up to $L D/2 + n$ to ensure that the integral is
projective.
\end{Theorem}

In the simple case where all the scalar products are squares of momenta, that is to
say, $s_0 = s_1$, all the terms in $U^{S_0}$ are equal and the mean is simply
$U^{S_0,S_0}$, which itself is the derivative of~$U$ with
respect to the variables associated to the edges in~$S_0$. Since a numerator of the
form $p_i\cdot p_i$ can also be obtained by lowering by one unit the exponent of
the propagator associated to edge~$i$, this allows for a simple proof of
Theorem~\ref{fund} in this case, using integration by parts in each of the Schwinger variables associated to the edges in
$S$. In fact, it is this
simple case which suggested that a symmetrical and rather simple result could be
obtained in the general case.

The interesting feature of Theorem~\ref{fund} is that, through the identity between the power of the graph
polynomial in the denominator and the argument of the $\Gamma$-function, it appears as a sum of the
evaluation of the graph in the dimensions $d+2$, $d+4$, up to $d+2m$, if $m$ is the total number of
scalar products in the numerator. The polynomial numerators mean that the propagators can become doubled,
tripled, or more, but since each factor is a kind of graph polynomial, of maximal degree 1 in each of its
variables, the maximal power of each propagator is $k+1$ for the term in dimension $d+2k$, meaning that we remain short of
having an infrared divergence on a single propagator. The precise interplay between this formulation and the structure of
subdivergences in this setting go beyond the purpose of this letter. This structure is similar to the one
appearing in the study of the graphs with subdivergences of~\cite{MaPaSch14}. As in this study, integration by part
identities should allow to relate terms stemming from different monomials in the numerator. On the other end, the explicit evaluation of terms with non trivial numerators in terms of higher dimensional graphs could decrease the total number of terms necessary to obtain a reduction to master integrals.

\section{Reducing to the completed diagram}
\subsection{A single scalar product}

In the case of a propagator graph $G$ in a scalar theory, it is known that the residue can be expressed as an integral which
only depends on the completed graph $\tilde G$ and its graph polynomial~$\tilde U$, with an exponent
for the additional edge giving scale invariance to the completed graph, which is then a vacuum graph (see, e.g.,~\cite{Sch10})
\begin{equation} \label{InS}
	\cI_{\tilde G} [\beta_i] = \frac {\Gamma(d/2)} {\prod_i \Gamma(\beta_i)}
			\int \prod_i \bigl( {\rm d}x_i^{\vphantom {\beta_i}} x_i^{\beta_i-1}\bigr)
			\tilde U^{-d/2} \delta \bigl(\textstyle{ \sum' }x_i -1\bigr).
\end{equation}
In both products, the index~$i$ goes over all the edges of $\tilde G$, while the sum $\sum'$ is over any non
empty subset.
The derivation of formula~\eqref{InS} uses the relation between the graph polynomial~$\tilde U$ of the completed
graph and the two Symanzik polynomials of the propagator graph: the first Symanzik polynomial is what have been called the graph
polynomial up to now and will be noted~$U$, while the second one, in this single scale setting, is $p^2 V$. The
deletion contraction relation then gives that $\tilde U = x_0 U + V$. It is traditional to give the
index~0 to the edge completing the graph.

In the case of a non trivial numerator, such an expression depending only on a completed graph seems harder to get by. Let us first study the case where the numerator is simply the scalar product $p_i\cdot p_j$ and redo the same steps as in the scalar case.
The first step is to integrate on the loop momenta. For fixed Schwinger parameters, it is a Gaussian integration with two
contributions for the scalar product, one coming from the correlation between $p_i$ and~$p_j$ which gets a factor of the dimension $d$, and the other stemming from the part proportional to the exterior momentum in both objects.
This latter part can be determined by the electric circuit analogy, where the $x_i$ parameters play the roles of
the resistance of the edges. The fraction of the intensity going through a given edge is given by the ratio of two
polynomials of degree $L$, one dependent on the edge $V^i$ and the universal $U$. By this analogy, for given
values of the parameters $x_i$, the mean of the momentum $p_i$ is given by $ p V^i/U$.

The evaluation of the diagram is then given, according to~\cite{Na71}, by
\[%\label{naive}
	\cI_G(p) = \frac {1}{ \prod_k \Gamma(\beta_k)} \int \prod_k \bigl( {\rm d}x_k^{\vphantom {\beta_k}} x_k^{\beta_k-1}\bigr)
 		\left(\frac d2 \frac{U^{i,j}}{U^{d/2+1}} + \frac{V^i V^j p^2}{U^{d/2+2}} \right) \exp\left(-p^2 \frac V U\right).
\]
The two terms do not have the same homogeneity degrees, since the $V^k$ have the same degree as $U$, while $U^{i,j}$ has one degree less.
Let us define $\beta_0$ such that $\beta_0 + \sum_i \beta_i = (L+1)d/2 +1$, so that the first term is homogeneous of degree $d/2 - \beta_0$
and the second is homogeneous of degree~${d/2 - \beta_0 +1}$ and integrate on the global scale of the Schwinger parameters. The $p^2$ already present in the second term makes the two terms proportional to \smash{$\bigl(p^2\bigr)^{\beta_0-d/2}$} so that the integrand without the $\Gamma$ prefactors and powers of the $x_j$ reads
\[%\label{naive_d}
 		\frac {d}2 \frac {\Gamma(d/2-\beta_0) U^{i,j}}{U^{\beta_0+1}V^{d/2-\beta_0}} + \frac{ \Gamma(d/2-\beta_0+1)V^i V^j}{U^{\beta_0+1}V^{d/2-\beta_0+1}} .
\]
We write $d/2$ in the first term as $\beta_0 + (d/2 - \beta_0)$ and use the functional identity for the $\Gamma$--function to obtain $\beta_0 \Gamma(d/2 - \beta_0) + \Gamma(d/2 - \beta_0 +1)$. The term with the $\beta_0$ factor can be written as an integral on $x_0$ as in the scalar case, giving
\begin{equation} \label{beta+1}
	\int_{x_0 =0}^\infty {\rm d}x_0 ^{\vphantom{\beta_0}} x_0 ^{\beta_0} \frac { \beta_0 \Gamma(d/2 + 1) }{\Gamma(\beta_0+1) } \frac { U^{i,j} } { \tilde U ^{d/2 +1} }.
\end{equation}
The $\beta_0$ factor of the numerator can be used to change the argument of the $\Gamma$--function in the denominator to $\beta_0$.

The other term has a denominator with a global degree in $U$ and $V$ of $d/2 +2$ and a simple application of the same transformation would give a higher power of $\tilde U$ in the denominator. We obtain indeed
\begin{equation}\label{intermediaire}
	\Gamma(d/2 - \beta_0 +1) \frac { V U^{i,j} + V^i V^j } { U^{\beta_0+1} V^{d/2 - \beta_0 +1}}
\end{equation}
by putting everything on the same denominator. However, the Dodgson identities allow to simplify the numerator. In the completed graph, we have
\begin{equation} \label{dodV}
	\tilde U \tilde U^{0i,0j} = \tilde U^{0,0} \tilde U^{i,j} - \tilde U^{0,j} \tilde U ^{i,0}.
\end{equation}
The part proportional to $x_0$ is trivially true but the part independent of~$x_0$ is quite interesting. It is obtained by adding the subscript 0
to the polynomials which do not have it in their upper indices. The definition of $\tilde U$ gives us that $\tilde U^{0,0} = U$
and~$\tilde U_0 = V$. The polynomial $V^k$ is defined by considering the trees in~$G$ including the edge $k$, with a sign dependent on the
orientation of~$k$ with respect to the path going from the entrance to the exit of the diagram~$G$. In the case of $\tilde U^{0,k}$, we must
consider subgraphs with a unique loop including both $0$ and~$k$, and a~sign is added if the two do not have the same orientation with respect to
this loop. The two sets of subgraphs are clearly in correspondence, with matching determination of the signs, so that we have that $\tilde U^{0,k}
= \tilde U^{k,0} = V^k$. The part for $x_0=0$ of equation~\eqref{dodV} can therefore be written~$V U^{i,j} = U \tilde U^{i,j}_0 - V^i V^j$.
The numerator in equation~\eqref{intermediaire} turns out to be proportional to~$U$, reducing the total power of $U$ and
$V$ in the denominator. The trick of writing this term as an integral over $x_0$ will give a~contribution similar to the one in equation~\eqref{beta+1}, so that both combine to give
\[%\label {fin}
	\int_{x_0 =0}^\infty {\rm d}x_0 ^{\vphantom{\beta_0}} x_0 ^{\beta_0-1} \frac { \Gamma(d/2 + 1) }{\Gamma(\beta_0)} \frac { x_0 U^{i,j} + \tilde U^{i,j}_0 } { \tilde U^{d/2+1} },
\]
where it is easy to recognize the full $\tilde U^{i,j}$ in the numerator.

\subsection{The general case}	\label{reducgene}

In the general case, such a step by step reduction of the terms coming from the parts proportional to the
exterior momentum would become rapidly very complex, without a clear generalization for an arbitrary
numerator. Indeed, when a numerator is a product, one must generate independently all possible powers of
$x_0$, splitting some terms in numerous parts. It is therefore impossible to separate a phase where one
reduces to the vacuum graph and one where the correlations are expressed in terms of the polynomials $U^S$.

However, it is possible to circumvent such a painful computation. We just start from the remark that
in a massless theory, the dependence on the exterior momentum is fixed by homogeneity reason to be a
power of $p^2$ which will be written $\beta_0 - d/2$. If we integrate $\exp\bigl(-p^2\bigr) \cI_G(p)$ on the
whole momentum space, one gets \smash{$ \Gamma(\beta_0)\pi^{d/2}/\Gamma(d/2) $} times the residue of the
graph, with the $d$ dependent factor simply half the volume of the $(d-1)$-sphere. This is linked to the use of~$p^2$ as the radial variable.
On the other hand, we could do the integration over this exterior~$p$ before all parametric
integrations, alongside the integrations on the momenta on the edges of the graph. These integration steps are just, for fixed Schwinger
parameters, the integration on the
completed graph $\tilde G$.
The parameter $x_0$ associated to the additional edge is simply fixed to~1. This constraint on $x_0$ can be
seen as a special case of the reduction of a projective integral if we complete the integrand
by a factor that makes it scale invariant. The relation between the original diagram $G$ and the completed one~$\tilde
G$ with one more loop shows that the required factor is $x_0^{\beta_0-1}$. The $\Gamma$ factors in the relation
between the residue of~$G$ and the integral on~$\tilde G$ ensure that all edges of the completed diagram are associated
with integrals with the same factor $1/\Gamma(\beta_i)$. The case of the propagator
diagrams is therefore reduced to the one of vacuum diagrams.

A similar approach is also possible in the case of non-scalar propagators. Lorentz invariance and
the presence of a single available vector give a limited number of tensor structures. In the case of a vector propagator, the only two possible
Lorentz invariant tensor structure are~$g_{\mu\nu}$ and~$p_\mu p_\nu$. The completion to a vacuum graph using these same two structures
would give enough information from these two vacuum graphs to fully determine the propagator.

The case of massive theories is less clear cut. The scale invariance which allowed this simple derivation is no longer available, but it would
nevertheless be useful to reduce the maximal power in the denominator. The step by step derivation of the previous subsection would be however
difficult to generalise: I would rather bet on careful expansion of an analog of our final formula to prove that they allow to recover Nakanishi's
results from a simpler expression. This would however require its own exposition space.

\section{Proof of the general result}
The proof of Theorem~\ref{fund} is based on the usual expression of the integral of polynomials under a~Gaussian measure as a sum of products of elementary
correlators. In a first step, the different terms are grouped in a slightly redundant manner involving permutations of the $n$ scalar product in
the set~$S$. The second step show that the same expression is obtained from the formula in Theorem~\ref{fund} through the use of Dodgson identities and
simple combinatorial identities.

\subsection{The terms to find}

From the preceding section, only correlations in the numerator have to be considered, since the studied cases
can be reduced to vacuum diagrams. The parametric method consists in replacing the propagators by integrals on parameters of Gaussian functions of
the momenta. Mix in additional variables to write the conservation of momentum at each vertex as just an additional quadratic term in the
exponential, and for fixed Schwinger parameters, a Gaussian integral has to be evaluated. In the absence of numerators and for vacuum diagrams, one
just obtains a~power of the determinant of the quadratic form in the Gaussian, which is just the first Symanzik polynomial. The numerators envisaged
in Theorem~\ref{fund} are just polynomial of degree~$2n$ in the momenta, part of the variables in the Gaussian integral. It is well known that the integral
of such polynomials reduce to a sum of products of $n$ elementary correlations of the Gaussian measure and that there are
$ (2n-1)!! $ terms in the sum, corresponding to the possible groupings by pairs of the $2n$ factors in the polynomial.\footnote{Our main interest
is with the case where all these vectors are associated to different edges of the graph, but having many to one relations between the vectors and the
edges of the graph does not change our proof. However, writing explicitly this function would make these already cumbersome formulas illegible.}
If we name $\cal T$ the
set of possible pairings~$T$ of the~$2n$ vectors in the scalar products, the numerator multiplies the gaussian integral by the following factor
\begin{equation} \label{numerat}
	\sum_{T \in \cal T} \prod_{\{i,j\} \in S} g_{\mu_i,\mu_j} \prod_{\{k,l\} \in T} g^{\mu_k,\mu_l} \frac{U^{k,l}}{2U}.
\end{equation}
In each term, the $n$ factors of the metric in the product over $S$ and the $n$ factors of its inverse in the product over $T$ combine to
only produce a power of the dimension~$d$. In order to prove Theorem~\ref{fund}, the different partitions $T$ in $\cal T$ must be collected in groups
giving the same power of~$d$. The power of $d$ is the number of independent chains of contracted indices between alternating metrics (coming
from~$S$) and inverse metrics (coming from~$T$).

The terms in Theorem~\ref{fund} are characterized by a permutation $\sigma$ of $S$ and the couple of functions~$(\varepsilon, \tilde\varepsilon)$,
as will be precised in the next subsection. Going from such data to a partition in pairs $T$ can be done by taking the pairs in~$T$ to be of the
form $\{\varepsilon(i), \tilde\varepsilon(\sigma(i)) \} $. It is clear that this produces a partition of the $2n$ vectors in $n$ pairs. In the
reverse direction, the permutation can be built cycle by cycle. Starting from a pair in~$T$ which is not already covered, we must choose the
element which will be $\varepsilon(i)$. The second element of the pair then determines what is~$\sigma(j)$ as well as the value
of~$\tilde\varepsilon$ at the point~$\sigma(j)$ if it is different from~$j$. Starting from~$\varepsilon(\sigma(j))$, the same reasoning
allows to determine successively the $\sigma^k(j)$ until one comes back to~$j$.

For a given term, the permutation $\sigma$ as well as the choice of functions $(\varepsilon,\tilde
\varepsilon)$ are not unique: in the case where $\sigma$ is the identity permutation, the product will
not depend on $(\varepsilon,\tilde \varepsilon)$ and in any case, exchanging the two functions
$(\varepsilon,\tilde \varepsilon)$ and changing $\sigma$ to its inverse leave the term unchanged. More
generally, the number of different choices of the function $\varepsilon$ compatible with a given term is
equal to $2^c$, where $c$ is the number of cycles of the permutation $\sigma$. This can be seen from the derivation of $\sigma$ from~$T$,
where we have a binary choice to make for each cycle.

When taking the average over the $2^n$ possible values of the pair $(\varepsilon,\tilde \varepsilon)$, this gives the product of the 2 in each of
the denominators for the correlations in equation~\eqref{numerat}. However, in this scheme each terms is produced multiple times and must be divided
by~$2^c$ to compensate if the permutation $\sigma$ has $c$ cycles. This number of cycles
also appears in the dependence on the dimension: inverse metrics from the correlations and metrics from the scalar products combine to give the trace of
the identity on the momentum space for each cycle of~$\sigma$, giving a factor~$d^c$.
Combining these two factors, the terms with a permutation of $c$ cycles must be multiplied by~$\bigl(\frac d 2 \bigr)^c$.
The signature of a permutation has also a simple expression in terms of the number of cycles, it is simply
the parity of $n-c$, with $n$ the number of objects on which the permutation acts.

\subsection[The terms in U\^{}S\_0]{The terms in $\boldsymbol{U^{S_0}}$}

We now start from the expression appearing in Theorem~\ref{fund}. Since all terms are subject to the
same averaging on all the possible values of the $\varepsilon$ function, we can just forget about the
different maps $\varepsilon$ and consider a fixed one. If we put every term on the same denominator
$U^{d/2+ \#(S)}$, each $U^{\varepsilon(S_i),\tilde\varepsilon(S_i)}$ can receive a factor $U^{\#(S_i) - 1}$
and be converted by the Dodgson identities
into the determinant of a matrix with correlators $U^{i,j}$ as elements.

If we now expand each of these determinants in terms of permutations, we obtain terms which are entirely
similar to the ones obtained in the preceding section. A given permutation of the whole set~$S$ can be
obtained in a number of different ways according to its cycle structure. Indeed, a given cycle structure can
only be obtained from the product of determinants associated to a partition of~$S$ in subsets which are
union of (the sets subjacent to) cycles. The identity is for example compatible with any partition of~$S$,
while a permutation with only one cycle is only compatible with the trivial partition with only one subset.

For a given permutation, the signs coming from the various determinants always combine to the signature of the complete
permutation, since the signature is multiplicative. It is given by the
parity of the size of~$S$ minus the number of cycles. Combined with the sign in our formulas, we obtain
that the unique term with the finest partition compatible with the cycle structure will have a positive sign
and the factor $\Gamma(d/2 + c)$, with $c$ the number of cycles in the permutation. We therefore will
have contributions from partitions of~$S$ with~$c$ down to $1$~parts, which comes with alternating
signs. All terms with the same number of parts in the partition are identical. Their numbers are
given by the Stirling numbers of the second kind $\left\{\begin{smallmatrix} c \\ p \end{smallmatrix} \right\}$, which
count the number of ways the $c$ cycles of the permutation can be collected in $p$
parts.\footnote{The notations used are the one proposed by Donald Knuth
in~\cite{Knuth_1992}. They have the advantage of circumventing the notation clash between combinatorists and analysts
which use the notation $(x)_n$ for respectively falling and raising factorials.}
We therefore obtain the following coefficients for the terms with $c$ cycles
\begin{equation}
\label{sumOfPartitions}
	\sum_{p=1}^c (-1)^{c-p}
	\left\{ \begin{matrix} c \\ p \end{matrix} \right\}
	\Gamma(d/2 + p).
\end{equation}
Now $\Gamma(d/2 + p)$ can be converted, using the functional equation for $\Gamma$, in the product of
$\Gamma(d/2)$ and the raising factorial or Pochhammer symbol $ (d/2)^{\overline p}$. It is a well-known property of the
Stirling numbers of the second kind that they allow to convert from falling factorials to plain powers, as it was
the reason for their introduction by Stirling in~\cite{Stirling_1730}, or by a
simple change of signs, from raising factorials to plain powers: all we need is the product rule
\[x \times
x^{\overline p} = x^{\overline{p+1}} - p x^{\overline p}
\]
 and the corresponding one for $x \times
x^{\underline p}$ and the triangle property of the Stirling number to obtain that the sum in
equation~\eqref{sumOfPartitions} will give $(d/2)^c \Gamma(d/2)$. This completes our proof of Theorem~\ref{fund}.

\section{Fermion loops}
Fermion loops could simply be reduced to a sum of terms of the kind just studied, but this would
completely defeat the interest of the method, which is to reduce the number of different polynomial factors
we have to compute, particularly the ones with the highest degree corresponding to a single correlation
$U^{i,j}$. We limit ourselves to the massless case, to be able to still use the method of Section~\ref{reducgene} to only
consider vacuum diagrams, with the added benefit that only loops with even number of propagators have to be considered in
the Yukawa case.
It is however interesting to keep the feature that the dimension of space only appears in the
$\Gamma(d/2+k)$ factor in the term with the power $d/2+k$ of the Symanzik polynomial in the denominator.
I therefore start from a set of correlations between numerators in the fermion loop and evaluate
the dimension depending traces coming from the contractions of gamma-matrices with the metric
associated to the correlation. Sums of products of correlations stemming
from the Dodgson identities are then used to obtain a nicer presentation for the result. As the number of terms to consider
grows rapidly with the number of terms in the loop, we shall present only the case of the loops of
four and six propagators, with only Yukawa couplings. Gauge couplings rapidly lead to more
cumbersome situations and will not be dealt with here.

\subsection{Loop of four fermionic propagators} \label{fourdirac}

In a simple 4 propagator loop, $\Tr ( \dirp_a \dirp_b \dirp_c \dirp_d )$, we obtain three possible
terms corresponding to the possible positions of the correlations, that is $U^{a,b} U^{c,d}$,
$U^{b,c} U^{d,a}$ and $U^{a,c} U^{b,d}$. The metric factor in the correlations get contracted with the Dirac matrices in the loop, giving
respectively the factors~$\Tr (\gamma_\mu \gamma_\mu \gamma_\nu \gamma_\nu)$,
 $\Tr (\gamma_\nu \gamma_\mu \gamma_\mu \gamma_\nu)$ and
 $\Tr (\gamma_\mu \gamma_\nu \gamma_\mu \gamma_\nu)$. They evaluate easily to $d^2 \Tr (1)$ for the first
two and $d(2-d) \Tr(1)$ for the last. The Dodgson identities then allow to write
$U^{a,b} U^{c,d} - U^{a,c} U^{b,d}$ as $U U^{ad,bc} $ and $U^{b,c} U^{d,a} - U^{a,c} U^{b,d}$ as $U U^{ba,cd}$.
The question then is which product will be taken and which ones will be eliminated with these
relations. The product $U^{a,c} U^{b,d}$ would seem to be nicer, since it does not break the
symmetries of the loop, however, depending on correlations of non-adjacent momenta, it may involve
terms with different signs when the correlations of adjacent momenta are merely obtained by
differentiations of the polynomial $U$. Without the overall factor $\Tr(1)$, we can make the
following transformations of the numerator associated to this fermion loop, including the one half factors in the
correlators which were not included previously
\begin{align}
 	N &= \frac{d^2}4 U^{a,b} U^{c,d} + \frac {d^2}4 U^{b,c} U^{d,a} +\frac {d (2-d)}4 U^{a,c} U^{b,d} \nonumber \\
	&= \frac{d^2}4 U^{a,b} U^{c,d} +\frac{d^2}4 U U^{ba,cd} + \frac d2 U^{a,c} U^{b,d} \nonumber \\
	&= \frac d2 \left( \frac d2 + 1\right) U^{a,b} U^{c,d} + \frac{d^2}4 U U^{ba,cd} - \frac d2 U U^{ad,bc} \nonumber \\
	&= \frac d2 \left( \frac d2 + 1\right) \bigl( U^{a,b} U^{c,d} + U U^{ba,cd} \bigr)
		- \frac d2 U \bigl( U^{ba,cd} + U^{ad,bc} \bigr). \label{4fin}
\end{align}
In this form, we obtain readily the factors necessary to convert $\Gamma(d/2)$ to $\Gamma(d/2+1)$
or $\Gamma(d/2+2)$. Separating the term proportional to $U U^{ba,cd}$ in two parts, so that the
dependence on $d$ is purely through the $\Gamma$ coefficients, appears to be necessary to obtain
nice results. In the next subsection, we will show the nice properties of the numerators coming from this choice.
Different forms of the numerator can be obtained if
one chooses to use different Dodgson identities, but insisting that terms quadratic in $d$ always
come as part of $d(d+2)$ makes the remaining terms proportional to $d$ independent of these
choices.

It is however interesting to further express the polynomials $U^{ad,bc}$ in terms of the spanning
forest polynomials introduced by Brown and Yeats~\cite{BrYe09}. These polynomials have the advantage
of having all their coefficients positive. They are no longer defined by sets of edges but by a~partition of the subset of vertices of the graph which are endpoints of the edges in $I\cup J$.
Then the forest polynomials, as their names indicate, are defined from forests with the same number of components as the
partition used to define them. If we have a partition with $k$~components, we consider the set of covering $k$-trees
of $G\backslash(I\cup J)$ such that all elements of a part of the partition belong to a corresponding tree in the $k$-tree. To
each such $k$-tree we associate a term in the forest polynomial, the product of the variables associated to the edges
{\em not} in the $k$-tree. A given forest polynomial associated with a partition in $k$~parts will contribute to a Dodgson
polynomial~$U^{I ,J}$ if $I$ and~$J$ have
$k-1$ elements and if the elements of~$I$ provide links between the different part of the partition. The same must be true for
the elements of~$J$. This means that both $I$ and~$J$ can complete the $k$-trees appearing in the definition of the forest
polynomial into covering trees.

In the particular
case of $U^{ad,bc}$, we need to consider the set of vertices $\{0,1,2,3\}$ such that $a$ is the
edge $(0,1)$ up to $d$ being $(3,0)$. In this case, each of the $U^{ad,bc}$ polynomials
corresponds to a unique spanning forest polynomial, but for the six-propagator case, it will be more
natural to express the numerators in terms of these `vertex' polynomials, combining the coefficients
of different `edge' polynomials.

Indeed, since there are two elements in either of the sets $I$ and $J$, the vertices must be in
three distinct sets, meaning that the possible polynomials have just a pairing of two vertices, the
other two being independent. We obtain that
\[
	U^{ad,bc} = U^{\{0,2\},1,3},\qquad
	U^{ba,cd} = U^{\{1,3\},0,2}.
\]
Our notation for the forest polynomial is that we have as indices the subsets in the partition, but we lighten it by
letting down the braces for the singletons.
This ends up the general properties of the four propagator fermion loop.

It is interesting to compare the result of equation~\eqref{4fin} to the application of Theorem~\ref{fund} to the expansion of the trace in terms of product of
scalar products, that is $ (p_a\cdot p_b) (p_c \cdot p_d) - (p_a \cdot p_c)(p_b \cdot p_d)+ (p_a \cdot p_d)(p_b \cdot p_c)$, again without an explicit
value for the trace of the identity in spinor space. The missing terms $U^{a,d} U^{b,c} - U^{a,c} U^{b,d}$ are simply the result of the expansion
of $U U^{ba.cd}$ with the Dodgson identities. Among the six terms involving a single polynomial, straight forward cancellation give the unique
term $U^{ac,bd}$. However, the three polynomials corresponding to the three different ways of splitting the four edges of the fermion loop in two
groups of two are not independent, so that we recover the result in equation~\eqref{4fin}. The relation is true for any choice of the edges, but is
easily seen in the present case by comparing the expansion in terms of forest polynomials of the three polynomials. In the more complex case I will
present in the following subsections, such comparison would be much harder, since the relations which can be used as well as the number of different
terms is much larger. The problem is particularly acute in the case of the intermediate dimension terms: we would need to expand products of the
polynomial of degree $L$ with polynomials of degree $L-3$ in terms of products of polynomials of degree~${L-1}$ times polynomials of degree
$L-2$, but there are numerous way in which it can be done. Collecting the 90 possible products to form the much lower number of terms obtained in
the following would require a disproportionate effort.

\subsection{Dimensional factors for six fermionic propagators} \label{sixdirac}

This case is more complex, since we now have fifteen different terms. The same sets of pairings of edges can than be
interpreted either as giving products of scalar products or as sets of correlations. And in the end,
we should produce three different numerators for the terms with denominators the Symanzik polynomial to the
powers $d/2+1$, $d/2+2$ and~$d/2+3$, interpretable as scalar integrals in dimensions $d+2$, $d+4$
and~$d+6$. Things can be made manageable if one considers the symmetries by translation along the loop. This
may not be a symmetry for the whole diagram, but it will collect terms with similar factors, obtained from
similar functions. The fifteen possible terms are grouped in five collections, with one, two, two times three
and six pairing schemes. It is helpful to have a graphical description of the possible pairing schemes. The
edges of the fermion loop are represented by the summits of an hexagon and the correlations are indicated by
lines. The five correlation schemes are therefore represented as
\begin{gather*}% \label{6corr}
a) \begin{array}{c}
\begin{tikzpicture}[scale=0.4]
	\draw (-1,0) -- (1,0); \draw (-0.5,-1) -- (0.5,1); \draw (-0.5,1) -- (0.5,-1);
\end{tikzpicture} \\ [4 ex]
\mkern -10 mu U^{a,d} U^{b,e} U^{c,f} \end{array}
b)\begin{array}{c} \begin{tikzpicture}[scale=0.4]
	\draw (-1,0) -- (-0.5,1); \draw (-0.5,-1) -- (0.5,-1); \draw (0.5,1) -- (1,0);
\end{tikzpicture} \\ [4 ex]
\mkern -10 mu U^{a,b} U^{c,d} U^{e,f} \end{array}
c)\begin{array}{c} \begin{tikzpicture}[scale=0.4]
	\draw (-1,0) -- (1,0); \draw (-0.5,-1) -- (0.5,-1); \draw (-0.5,1) -- (0.5,1);
\end{tikzpicture} \\ [4 ex]
\mkern -10 mu U^{a,d} U^{b,c} U^{e,f} \end{array}
d)\begin{array}{c} \begin{tikzpicture}[scale=0.4]
	\draw (-1,0) -- (1,0); \draw (-0.5,-1) -- (-0.5,1); \draw (0.5,-1) -- (0.5,1);
\end{tikzpicture} \\ [4 ex]
\mkern -10 mu U^{a,d} U^{b,f} U^{c,e} \end{array}
e)\begin{array}{c} \begin{tikzpicture}[scale=0.4]
	\draw (-1,0) -- (0.5,1); \draw (-0.5,-1) -- (0.5,-1); \draw (-0.5,1) -- (1,0);
\end{tikzpicture} \\ [4 ex]
\mkern -10 mu U^{a,c} U^{b,d} U^{e,f} \end{array}
\end{gather*}
with the corresponding algebraic expression beneath them.
It is easy to see that case $a)$ is invariant by rotation, case $b)$ has two versions, cases $c)$
and~$d)$ three and that for $e)$, all six possible rotations are distinct. We also need the associated
dimensional factors, when taking the trace of the product of six $\gamma$ matrices with the given pattern of
contractions. All results are given without the factor of the trace of the identity in spinor space.
In cases $b)$ and~$c)$, the factor is simply $d^3$, cases $d)$ and~$e)$
only use the relation $\gamma_\mu \gamma_\rho \gamma^\mu = (2-d) \gamma_\rho$ to obtain the respective factors
$d(2-d)^2$ and~$d^2(2-d)$. The only factor remaining to compute is for $a)$ and requires some work with
the anticommutation relations of Dirac matrices to obtain $-d\bigl(d^2-6d+4\bigr)$. It is remarkable that the
same results can be obtained from the requirement that the sum of all coefficients be $d(d+2)(d+4)$, similarly
to the four propagator case where the sum of the coefficients is $2 d^2 + d(2-d) = d(d+2)$. This relation
ensures that, after using Dogson identities to keep only one product of three correlations, its coefficient will
be proportional to $d(d+2)(d+4)$, suitable to be interpreted as the $\Gamma$ factor for a $(d+6)$-dimensional
integral. Remembering that each correlator comes with an $1/2$ factor, the different correlators will come
with the following dimensionally dependent factors, expressed in terms of the raising factorials of $d/2$
\begin{gather*}
	 a) - \left(\frac d 2 \right)^{\overline 3} + 6 \left(\frac d 2 \right)^{\overline 2}
		- 5 \left(\frac d 2 \right)^{\overline 1} ,\qquad
	 b),c) + \left(\frac d 2 \right)^{\overline 3} - 3 \left(\frac d 2 \right)^{\overline 2}
		+ \hphantom{1} \left(\frac d 2 \right)^{\overline 1} ,\\
	 d) + \left(\frac d 2 \right)^{\overline 3} - 5 \left(\frac d 2 \right)^{\overline 2}
		+ 4 \left(\frac d 2 \right)^{\overline 1} ,\qquad
	 e) - \left(\frac d 2 \right)^{\overline 3} + 4 \left(\frac d 2 \right)^{\overline 2}
		- 2 \left(\frac d 2 \right)^{\overline 1}.
\end{gather*}
These presentations in terms of the raising factorials will guide our search for the expression of the
numerators, with the terms proportional to $d/2$ the most constrained, since they must be expressed as sums of
simple Dodgson polynomials.

\subsection{Dodgson polynomials for the six propagator loop}

As in the case of the loops of four propagators, we will take as reference situations the ones with only nearest
neighbour correlations, i.e., configurations of type~$b)$. Configurations $c)$ and~$e)$ have a nearest neighbour
correlation, so that they can be obtained from a product of this nearest neighbour correlator and a simple Dodgson
polynomial involving only four of the propagators of the loop, similar to the ones appearing in the preceding
subsection. The new objects we must consider therefore have to produce configurations $a)$ and~$d)$. They
will involve Dodgson polynomials with sets $I$ and~$J$ forming a partition of the six edges in the fermionic
loop. There are ten different partitions, but three of them have no interest for us: they are the ones with
contiguous parts (of the model $[abc][def]$) which do not allow for three nearest neighbour links between the two
parts. The seven others are divided between a highly symmetric one, $[ace][bdf]$, represented graphically in
equation~\eqref{sympart}, which produces, inter alia,
configuration~$a)$ and the six rotations of $[ade][bcf]$, represented in equation~\eqref{oddpart},
which can produce the configurations of set~$d)$. The order of the indices in the corresponding Dodgson polynomials, which could
introduce an additional sign, is fixed through the fact they are in the same order along the loop in $I$ and $J$: this fixes the order up to a
cyclic permutation which is even.
The lower number of configurations in set~$d)$ with respect to the partitions of this subset means that some
latitudes remain in our choice of Dodgson polynomials generating the required terms.\looseness=1

It will be convenient to have a graphical representation of the different terms coming from a~given Dodgson
polynomial. The edges from one set will be marked by a white blob and we will show the correlators coming from
the terms of the determinant associated to each permutation. We obtain for these two kinds of partitions the
following set of correlators, with the terms aligned according to the permutation they come from
\begin{gather}
\begin{tikzpicture}[scale=0.2]
 \node (g) at (-2,0)[bla]{}	; \node (hg) at (-1,2)[noi]{}	; \node (bg) at (-1,-2)[noi]{}	;
 \node (hd) at (1,2)[bla]{}	; \node (bd) at (1,-2)[bla]{}	; \node (d) at (2,0)[noi]{}	;
\end{tikzpicture}
=
\begin{tikzpicture}[scale=0.2]
 \node (g) at (-2,0)[bla]{}	; \node (hg) at (-1,2)[noi]{}	; \node (bg) at (-1,-2)[noi]{}	;
 \node (hd) at (1,2)[bla]{}	; \node (bd) at (1,-2)[bla]{}	; \node (d) at (2,0)[noi]{}	;
 \draw (g) to (hg); \draw (hd) to (d); \draw (bd) to (bg) ;
\end{tikzpicture}
 +
\begin{tikzpicture}[scale=0.2]
 \node (g) at (-2,0)[bla]{}	; \node (hg) at (-1,2)[noi]{}	; \node (bg) at (-1,-2)[noi]{}	;
 \node (hd) at (1,2)[bla]{}	; \node (bd) at (1,-2)[bla]{}	; \node (d) at (2,0)[noi]{}	;
 \draw (g) to (d); \draw (hd) to (bg); \draw (bd) to (hg) ;
\end{tikzpicture}
 +
\begin{tikzpicture}[scale=0.2]
 \node (g) at (-2,0)[bla]{}	; \node (hg) at (-1,2)[noi]{}	; \node (bg) at (-1,-2)[noi]{}	;
 \node (hd) at (1,2)[bla]{}	; \node (bd) at (1,-2)[bla]{}	; \node (d) at (2,0)[noi]{}	;
 \draw (g) to (bg); \draw (hd) to (hg); \draw (bd) to (d) ;
\end{tikzpicture}
 -
\begin{tikzpicture}[scale=0.2]
 \node (g) at (-2,0)[bla]{}	; \node (hg) at (-1,2)[noi]{}	; \node (bg) at (-1,-2)[noi]{}	;
 \node (hd) at (1,2)[bla]{}	; \node (bd) at (1,-2)[bla]{}	; \node (d) at (2,0)[noi]{}	;
 \draw (g) to (hg); \draw (hd) to (bg); \draw (bd) to (d) ;
\end{tikzpicture}
 -
\begin{tikzpicture}[scale=0.2]
 \node (g) at (-2,0)[bla]{}	; \node (hg) at (-1,2)[noi]{}	; \node (bg) at (-1,-2)[noi]{}	;
 \node (hd) at (1,2)[bla]{}	; \node (bd) at (1,-2)[bla]{}	; \node (d) at (2,0)[noi]{}	;
 \draw (g) to (bg); \draw (hd) to (d); \draw (bd) to (hg) ;
\end{tikzpicture}
 -
\begin{tikzpicture}[scale=0.2]
 \node (g) at (-2,0)[bla]{}	; \node (hg) at (-1,2)[noi]{}	; \node (bg) at (-1,-2)[noi]{}	;
 \node (hd) at (1,2)[bla]{}	; \node (bd) at (1,-2)[bla]{}	; \node (d) at (2,0)[noi]{}	;
 \draw (g) to (d); \draw (hd) to (hg); \draw (bd) to (bg) ;
\end{tikzpicture} ,\label{sympart}
 \\[1.5ex]
\begin{tikzpicture}[scale=0.2]
 \node (g) at (-2,0)[bla]{}	; \node (hg) at (-1,2)[noi]{}	; \node (bg) at (-1,-2)[noi]{}	;
 \node (hd) at (1,2)[noi]{}	; \node (bd) at (1,-2)[bla]{}	; \node (d) at (2,0)[bla]{}	;
\end{tikzpicture}
=
\begin{tikzpicture}[scale=0.2]
 \node (g) at (-2,0)[bla]{}	; \node (hg) at (-1,2)[noi]{}	; \node (bg) at (-1,-2)[noi]{}	;
 \node (hd) at (1,2)[noi]{}	; \node (bd) at (1,-2)[bla]{}	; \node (d) at (2,0)[bla]{}	;
 \draw (g) to (hg); \draw (d) to (hd); \draw (bd) to (bg) ;
\end{tikzpicture}
 +
\begin{tikzpicture}[scale=0.2]
 \node (g) at (-2,0)[bla]{}	; \node (hg) at (-1,2)[noi]{}	; \node (bg) at (-1,-2)[noi]{}	;
 \node (hd) at (1,2)[noi]{}	; \node (bd) at (1,-2)[bla]{}	; \node (d) at (2,0)[bla]{}	;
 \draw (g) to (hd); \draw (d) to (bg); \draw (bd) to (hg) ;
\end{tikzpicture}
 +
\begin{tikzpicture}[scale=0.2]
 \node (g) at (-2,0)[bla]{}	; \node (hg) at (-1,2)[noi]{}	; \node (bg) at (-1,-2)[noi]{}	;
 \node (hd) at (1,2)[noi]{}	; \node (bd) at (1,-2)[bla]{}	; \node (d) at (2,0)[bla]{}	;
 \draw (g) to (bg); \draw (d) to (hg); \draw (bd) to (hd) ;
\end{tikzpicture}
 -
\begin{tikzpicture}[scale=0.2]
 \node (g) at (-2,0)[bla]{}	; \node (hg) at (-1,2)[noi]{}	; \node (bg) at (-1,-2)[noi]{}	;
 \node (hd) at (1,2)[noi]{}	; \node (bd) at (1,-2)[bla]{}	; \node (d) at (2,0)[bla]{}	;
 \draw (g) to (hg); \draw (d) to (bg); \draw (bd) to (hd) ;
\end{tikzpicture}
 -
\begin{tikzpicture}[scale=0.2]
 \node (g) at (-2,0)[bla]{}	; \node (hg) at (-1,2)[noi]{}	; \node (bg) at (-1,-2)[noi]{}	;
 \node (hd) at (1,2)[noi]{}	; \node (bd) at (1,-2)[bla]{}	; \node (d) at (2,0)[bla]{}	;
 \draw (g) to (bg); \draw (d) to (hd); \draw (bd) to (hg) ;
\end{tikzpicture}
 -
\begin{tikzpicture}[scale=0.2]
 \node (g) at (-2,0)[bla]{}	; \node (hg) at (-1,2)[noi]{}	; \node (bg) at (-1,-2)[noi]{}	;
 \node (hd) at (1,2)[noi]{}	; \node (bd) at (1,-2)[bla]{}	; \node (d) at (2,0)[bla]{}	;
 \draw (g) to (hd); \draw (d) to (hg); \draw (bd) to (bg) ;
\end{tikzpicture} .\label{oddpart}
\end{gather}
In the case of the symmetric partition, equation~\eqref{sympart} shows that one gets the correlation of
type~$a)$, the two ones of type~$b)$ as well as the three ones of type~$c)$ with a minus factor. The
fact that only full sets appear was predictable, since there is only one partition of this kind. In the case of
the other kind of partitions, we obtain one of the elements of group~$d)$ as well as the element of
group~$c)$ sharing the same diagonal correlator with a minus sign. We also have an element of group~$b)$
as well as three in group~$e)$, but two with a minus sign and one with a plus sign.\looseness=1

We are now in a position to express the terms with a factor $d/2$. For the correlation of type $a)$, which
comes with the coefficient~$-5$, we only need to take $-5$ times the Dodgson polynomial with the symmetric
partition. The coefficient~4 for the correlation of type~$d)$ will be obtained by adding two times each of the
six possible Dodgson polynomials of the other kind. We then obtain~4 times each of the correlations of
type~$d)$ as well as those of type~$c)$, but with a minus sign, since two of the six positions contribute to
each element in this case. For type~$c)$, combined with the 5 coming from the other type of polynomial,
one obtains the required factor 1. For type~$b)$, there are three positions which contribute to each one,
giving a factor~6 which combines with the $-5$ to give the required 1. Finally, each object of type~$e)$
will see one configuration giving a positive contribution and two a negative one, totalling the expected~$-2$
coefficient. We were therefore able to give all these contributions as linear combinations of Dodgson
polynomials. This was predictable from Theorem~\ref{fund} applied to the fifteen different products of scalar
products coming from the expansion of the trace, but such a derivation would be much more
cumbersome.

Next we have to consider the terms proportional to $(d/2)^{\bar 2}$. We certainly need 6 of the symmetric
Dodgson polynomials to get the proper number of objects of type $a)$. However, obtaining the $-5$ coefficient
for type~$d)$ is not so clear, since an equal distribution among the~6 Dodgson polynomials would introduce
half integers. As in
the four fermion propagator case, we will show that reducing the symmetry allows for a reduction of the number of
different terms. We therefore split this set between odd and even ones. For definiteness, say that the one
represented in equation~\eqref{oddpart} is an even one, as well as the two ones obtained by rotations by~2 or 4 units.
There is a corresponding split of sets $b)$ and~$e)$ into even and odd parts, with the even polynomial giving
the even element of set~$b)$, two even elements of set~$e)$ with a minus sign and one odd element of
set~$e)$ with a plus sign. Suppose now that we take $-3$ times the even partitions and~$-2$ times the odd
ones. For sets $c)$ and~$d)$, only the total number counts and we get $-5$ times each element of~$d)$ as
well as 5 times the elements of~$c)$, which combined with the~$-6$ from the symmetric partition gives~$-1$.
For the even element of~$b)$, we get $-9$ from the~3 even partitions, which combined with the 6 gives the
required $-3$. For the odd one, the odd partitions only give~$-6$ which cancels the other contribution. In
the case of the elements of~$e)$, we have the same pattern that we have the proper count for the even elements
and a missing factor for the odd ones. Indeed, the even elements receive a factor 6 from the even partition and
$-2$ from the odd ones, adding to 4, while for the odd elements, the factors are $-3$ and~4, adding to~1.

After using the Dodgson polynomials involving the six propagators of the loop, we therefore still have to account
for $-3$ times the odd element of ~$b)$, 3 times every odd elements of~$e)$ as well as $-2$~times the
elements of~$c)$. However, for these terms, we can use products of a~nearest neighbour correlator and a
simpler Dodgson polynomial with the sets $I$ and~$J$ having two elements. We can produce in this way the
difference between the odd element of~$b)$ and each of the odd elements of~$e)$ or the difference between an
element of~$c)$ and a corresponding odd element of~$e)$. One of the set of odd elements of~$e)$ will be
paired with the odd element of~$b)$, giving one of the three terms of type~B
\begin{equation}
\label{Bterm}
B\colon\ U^{a,b} U^{c,d} U^{e,f} - U^{a,b} U^{c,e} U^{d,f} = U U^{a,b} U^{cf,de}.
\end{equation}
The two other sets will fit with the elements of~$c)$, giving two times the terms of type~A, according to the scheme{\samepage
\begin{equation}
\label{Aterm}
A\colon\ U^{a,b} U^{c,f} U^{e,d} - U^{a,b} U^{c,e} U^{d,f} = U U^{a,b} U^{cd,fe}.
\end{equation}
This ends the reduction of the numerator in the $d+4$~term.}

The last point is to consider the terms with a $(d/2)^{\bar 3}$ factor. The $-1$ factor for the type~$a)$
term is easily accounted for according to our now usual way. The term of type~$d)$ asks for a subtler
treatment. If only even polynomials were used, we would end up with too large an imbalance between even and odd
terms for the sets $b)$ and $e)$. It is however possible to get all three elements of type~$d)$ with two
polynomials of even type and one of odd type. The even polynomial of type~$b)$ is obtained two times and with
the $-1$ coming with the type~$a)$ term, we have the proper number. For the odd one, the two contributions
from these Dodgson polynomials compensate and the balance is closed by the product of three correlators, giving
the only truly 10 dimensional term.
For the elements of type~$e)$, this choice of Dodgson polynomials does not produce a net contribution for the
odd ones and gives the required $-1$ factor for the even ones. As in the preceding case, the odd elements of
type~$e)$ and the elements of type~$c)$ can be combined to give the terms of type A. One can remark that
in the end, we only need to explicitly compute the three `odd'
correlators appearing in the odd element of type~$b)$, since only these `odd' correlators appear in the composite terms of types
A and~B.

\subsection{Forest polynomials}

Evaluation of Dodgson polynomials from their definitions as determinants remains challenging. The forest
polynomials suggested in~\cite{BrYe09} can however be evaluated from simple combinatorial rules, especially when
their polynomial degree remains low, as in the application we have here. The only delicate point is in the
determination of the signs which appear in the decomposition of the Dodgson polynomials, but they can be
determined once for all situations. The last point is that in our example, the contributions from different
Dodgson polynomials to each forest polynomial add to simple results.

The first step is to enumerate the possible forest structures. The six intermediate vertices in the cycle of
fermionic propagators must be grouped in four trees. There are therefore two possibilities, either three
isolated vertices and a tree grouping the three other ones or two isolated vertices and two pairs.
With the additional constraints that two neighbouring vertices cannot be isolated since no coloring of the three
adjacent edges allow them to be connected to the rest of the diagram with the two choices of colours, and that two
neighbouring vertices cannot be in the same tree since a loop would necessarily be made, we get the following four
types of forests
\begin{equation}
1)\begin{array}{c} \begin{tikzpicture}[scale=0.2]
 \node (h) at (0,2)[ver]{}	; \node (hg) at (-2,1)[ver]{}	; \node (bg) at (-2,-1)[ver]{}	;
 \node (hd) at (2,1)[ver]{}	; \node (bd) at (2,-1)[ver]{}	; \node (b) at (0,-2)[ver]{}	;
 \node (c) at (0,0) [circle, fill=black, minimum size=0.1mm] {};
 \draw (c) to (h); \draw (c) to (bg); \draw (c) to (bd);
\end{tikzpicture} \\[4ex]
U^{\{0,2,4\},1,3,5} \end{array} \
2) \begin{array}{c} \begin{tikzpicture}[scale=0.2]
 \node (h) at (0,2)[ver]{}	; \node (hg) at (-2,1)[ver]{}	; \node (bg) at (-2,-1)[ver]{}	;
 \node (hd) at (2,1)[ver]{}	; \node (bd) at (2,-1)[ver]{}	; \node (b) at (0,-2)[ver]{}	;
 \draw (b) to (h); \draw (bg) to (bd);
\end{tikzpicture} \\[4ex]
U^{\{0,4\},\{2,5\},1,3} \end{array}\
3) \begin{array}{c} \begin{tikzpicture}[scale=0.2]
 \node (h) at (0,2)[ver]{}	; \node (hg) at (-2,1)[ver]{}	; \node (bg) at (-2,-1)[ver]{}	;
 \node (hd) at (2,1)[ver]{}	; \node (bd) at (2,-1)[ver]{}	; \node (b) at (0,-2)[ver]{}	;
 \draw (h) to (b); \draw (hg) to (bd);
\end{tikzpicture} \\[4ex]
U^{\{1,4\},\{2,5\},0,3} \end{array}\
4) \begin{array}{c} \begin{tikzpicture}[scale=0.2]
 \node (h) at (0,2)[ver]{}	; \node (hg) at (-2,1)[ver]{}	; \node (bg) at (-2,-1)[ver]{}	;
 \node (hd) at (2,1)[ver]{}	; \node (bd) at (2,-1)[ver]{}	; \node (b) at (0,-2)[ver]{}	;
 \draw (h) to (bd); \draw (hg) to (b);
\end{tikzpicture} \\[4ex]
U^{\{1,5\},\{2,4\},0,3} \end{array}.\label{forests}
\end{equation}
All these configurations contribute to the symmetric Dodgson polynomial.
The configuration~$1)$ is the only possible one with three isolated points and has two versions, that we will
call even (the one represented) and odd. The even
one only contributes to the even polynomials and the odd one to the odd ones. The relative positions in
equations~\eqref{oddpart} and~\eqref{forests} are chosen so that the vertices and edges, all represented by dots,
alternate on a circle if we overlapped the figures.
When the two isolated vertices are at distance two, the only possibility for the other groups is given by the
configuration~$2)$, with six different possibilities. Here again, we can divide them into even ones and odd
ones. The one represented is an even one, and all even configurations contribute to all even polynomials but
not to the odd ones and vice versa. Finally, the configurations $3)$ and~$4)$ share diametrically opposed
isolated points, have three different versions each of which contributes to two polynomials, an odd one and an even one. The
pictured cases are the one contributing to the configuration appearing in equation~\eqref{oddpart}. The
only remaining point is the sign affecting all these forest polynomials. The situation is simple, since all
contributions come with a plus sign, except for the ones of type~$4)$ which have the minus sign, independently
on the Dodgson polynomial considered. The best way to compute these signs is to consider the diagram with the
six edges of the fermion loop and the trees of the forest as vertices. The signs then result from the
computation of a product of three by three determinants.\looseness=-1

With these relations between the Dodgson polynomials and the forest polynomials, it is easy to express the
results of the previous subsection in terms of forest polynomials. For types $3)$ and~$4)$, even and odd Dodgson
polynomials have the same contribution, so that the resulting coefficient is
just the sum of the coefficients for correlators of type~$a)$ and type~$d)$, with a~minus sign for
type~$4)$. We therefore have no contribution in the term with the factor~$(d/2)^{\bar 3}$, a~factor~$+1$
for type~$3)$ in the dimension $d + 4$ term and for type~$4)$ in the dimension $d+2$ term and a factor~$-1$ for the
remaining cases of~$3)$ in dimension $d+2$ and $4)$ in dimension $d+4$. For type $1)$ and~$2)$, the only relevant information is
whether they are even or odd and the total number of Dodgson polynomials of the same parity if we make the odd convention
that the symmetric Dodgson polynomial is both even and odd. This results in a factor 1 for all
these forest polynomials in the $d/2$ term, a
zero net contribution for the odd ones in the two other terms, while the even forest polynomials have
coefficients~$1$ in the $(d/2)^{\bar 3}$ term and~$-3$ in the $(d/2)^{\bar 2}$ one. The remarkable point is
that, at the level of these forest polynomials, the coefficients are notably simpler, only 0 and 1 in absolute
value, apart from the single $-3$.

This will be sufficient for the applications we present in this work, but
we must remember that this applies only to the case of a closed fermion loop. In some diagrams, we way have to
deal with a fermionic trace which is supported by an open path in the diagram, in which case new forest
polynomials could appear, where the two ends of the path do not belong to the same tree.\looseness=-1

Finally, the different coefficients are summed up in Table~\ref{summary}. The incomplete subsets used for the highest
dimensional terms are represented by $1/3$ and $2/3$, since the details of the used Dodgson polynomials have no
influence on the expression in terms of forest polynomials. Finally, A and B indicate the terms introduced in the previous
subsection which are products of a correlator and a Dodgson polynomial, introduced in equations \eqref{Aterm} and~\eqref{Bterm}.
We do not introduce the expression of the Dodgson polynomials appearing in these A and B terms as sum of forest polynomials, since none of them is
common to A and B terms. The single term in dimension
ten without a $U_G$ factor is not written as well as the $U_G$ factors in the other terms.

In order to make the application in Section~\ref{section6.3} smoother, I present also the obtained rational fraction to be integrated with respect to the
measure involving powers of the Schwinger parameters. It is written with a mix of Forest and Dodgson
polynomials. When three or six terms related by rotation appear, only one will be written and the symmetry must be applied to a~whole parenthesised
expression
\begin{align*}
{\mathcal I} ={}& \frac{\Gamma(d/2+1)}{U^{d/2+1}} \bigl( U^{\{0,2,4\},1,3,5} + U^{\{1,3,5\},0,2,4} + U^{\{0,4\},\{2,5\},1,3} + 5 \operatorname{sym} \nonumber \\
& - U^{\{0,4\},\{2,5\},1,3} - 2 \operatorname{sym} + U^{\{1,5\},\{2,4\},0,3} + 2 \operatorname{sym} \bigr) \nonumber\\
&-\frac{\Gamma(d/2+2)}{U^{d/2+2}} \bigl( U^{a,b} \bigl( 2 U^{cd,fe} + U^{cf,de} \bigr) + 2 \operatorname{sym}+ 3 U U^{\{0,2,4\},1,3,5} \nonumber \\
	&+ U \bigl(3 U^{\{0,4\},\{2,5\},1,3} - U^{\{1,4\},\{2,5\},0,3} + U^{\{1,5\},\{2,4\},0,3}\bigr) + 2 \operatorname{sym} \bigr) \nonumber \\
&+ \frac{\Gamma(d/2+3)}{U^{d/2+3}} \bigl( U^{a,b} U^{c,d} U^{e,f} + U^{a,b} U^{cd,fe} + 2 \operatorname{sym} \nonumber\\
& + U^2 U^{\{0,2,4\},1,3,5} + U^2 U^{\{0,4\},\{2,5\},1,3} + 2 \operatorname{sym} \bigr).
\end{align*}

\begin{table}[t]\centering\renewcommand{\arraystretch}{1.2}
\begin{tabular}{|l|c|c|c|c||c|c|c|c|c||c|c|c|c|}
\hline
&\multicolumn{4}{|c||}{Correlations} & \multicolumn{5}{|c||} {Dodgson} & \multicolumn{4}{|c|}{Forest}\\
\hline
&$a$& $b$, $c$& $d$& $e$& $s$&odd&even&A&B&1, 2 odd&1, 2 even&3&4\\
\hline
$d + 6$ & $-1$& $+1$ & $+1$ & $-1$ & $-1$ & 1/3 & 2/3 & 1 & 0 & 0 & 1 & 0 & 0 \\
$d+4$ & 6 & $-3$ & $-5$ & 4 & 6 & $-2$ & $-3$ & $-2$ & $-1$ & 0 & $-3$ & 1 & $-1$ \\
$d+2$ & $-5$& 1 & 4 & $-2$ & $-5$ & 2 & 2 & 0 & 0 & 1 & 1 & $-1$ & 1 \\
\hline
\end{tabular}
\caption{Recapitulation of factors for the loop of six fermions.}
\label{summary}
\end{table}

\section{Applications to the Wess--Zumino model}
\subsection{Supersymmetry and completion independence}

The importance of the supersymmetric Wess--Zumino model stems from the fact that it is the first supersymmetric model in
four spacetime dimensions to be found, before the advent of the supersymmetric extensions of gauge field theory or
gravity. Its field content is quite simple, with only scalar fields and fermion fields of the same mass and Yukawa
interactions between fermions and scalars and quartic self-interactions of the scalars. Supersymmetry implies that the
quartic scalar coupling is the square of the Yukawa coupling. The supersymmetric
transformation of the fermionic field involves a quadratic polynomial in the scalar fields and the algebra of
supersymmetric transformations is only verified up to the classical equations of motion. Both these facts make the study
of the quantum case difficult, but can be solved by the introduction of an auxiliary field. As the name implies, such a
field has classical equations of motion which do not allow for an independent propagation, since they constrain it to be
a quadratic polynomial of the scalars. The quartic coupling of scalars is decomposed in two cubic couplings of an
auxiliary field with two scalar field, each of the same strength as the Yukawa coupling.

One further advantage of this formulation with auxiliary fields is that the interaction vertex do not get any corrections:
the divergence in the quartic scalar interactions is simply one in the propagator of the auxiliary field, which is
therefore no longer constant.
To ensure supersymmetry, it is then sufficient that the ratios of the inverse propagators of all components of the
supermultiplet to their free counterparts be equal. This ratio is also the only necessary ingredient to compute the
renormalisation group functions, as explained in our previous works~\cite{BeSc08,BeSc12}. All considered graphs are
bipartite ternary graphs, but to a single topology, we associate a collection of
Feynman graphs with all possible particle assignments for each edge. The numerators we will compute give the sum of the
contributions for a given topology.

An important property of these numerators is that they only depend on the completed graph. The position of
the added edge in the completed graph does not matter. We only consider graphs without propagator
insertions, since only primitive graphs are studied,
while the Schwinger--Dyson equations take care of the decorations of the propagators.

In fact, both supersymmetry and independence from the added edge stem from the same fundamental graphical computation. Let
us consider the star with vertices $(a,b,c,d)$ and the edges $(a,b)$, $(a,c)$ and~$(a,d)$, such that $(a,b)$ is the scalar edge added to complete the graph for the auxiliary field propagator. All edges in this star are therefore of scalar type. Suppose now that we start from vertex $c$ and there is a auxiliary field edge $(c,f)$, which contribute a~factor~$p_{c,f}^2$. Using momentum conservation at the vertex $f$, $p_{c,f} = p_{f,g}+p_{f,h}$, one can write~\smash{$p_{c,f}^2 = \dirp_{c,f} \dirp_{f,g} + \dirp_{c,f} \dirp_{f,h}$}.
One can therefore replace the single auxiliary field numerator stemming from the auxiliary field on the edge $(c,f)$ by
the sum of two numerators stemming respectively from fermionic paths~$(c,f,g)$ and $(c,f,h)$. If the end of any of
these paths contacts an auxiliary field edge, the same procedure can be used to extend the path by two more edges in two
different ways. This extension procedure will end in one of two different ways: either one reaches vertex $b$, which is
the only vertex with only scalar incident edges in the completed graph which is of the good color ($a$ is unreachable),
or one encounters a fermionic line. This fermionic line can either be a~preexisting fermion loop or the line extending
from $c$. Any graph with a fermionic line extending from the vertex~$c$ is obtained once and only once through this construction, since the graph from which it is deduced can be recovered by replacing the line with alternating auxiliary field and bosonic field edges.

In the case of paths which do not extend to vertex $b$, the same set of fermionic edges can be
obtained in three different ways. Topologically, they form a stem (which can be empty) starting
form $b$, plus a loop: the loop can be either a preexisting fermion loop, or be covered in the two
possible directions. The product $\Pi$ of the propagators in the loop, being an even number of
terms proportional to $\gamma_\mu$, can be expressed on the basis of the even part of the Clifford
algebra
$\Pi = A \id + B^{\mu\nu} \gamma_{\mu\nu} + C \gamma_5$.
In this equation, $A$, $B^{\mu\nu}$ and~$C$ are functions of the momenta associated to the
edges in the loop. One can see that the term proportional to $\gamma_{\mu\nu}$ changes sign when
the loop is reversed and disappears in the sum over the two loops. When applying a chiral
projector, $\id$ and $\gamma_5$ become proportional, so that the sum of the two lassos is $2 A
\pm 2 C$ while the trace of the matrix on the loop becomes also $2A \pm 2C$. With the minus sign
associated to each fermion loop, these three contributions exactly cancel, independently on the
field assignments on the other edges of the graph.

We therefore obtain the same numerator when considering either all possible graphs with a~fermionic path from
$c$ to~$b$ or the ones without numerators on all neighbors of the edge $(a,b)$. We could obtain in a
similar way the sum of the graphs with a fermionic path from $c$ to~$b$ from the graphs which single
out the edge $(c,a)$. We therefore have that the numerators are the same if the completed graph was
obtained by adding the edge $(a,b)$ or the edge $(a,c)$.

Since we can move the singled out edge one step, it can be moved on any position by combining such steps. This
implies in particular the absence of vertex subdivergences, since by placing the singled out edge in a vertex
subgraph, it becomes ultraviolet convergent by a simple dimensional argument.

Supersymmetry is also a simple consequence. The fermionic two-point function has a~fer\-mion\-ic path from $a$
to $b$. It can either begin by the edge $(a,c)$, which is completed by a~path from $c$ to $b$, or
$(a,d)$, completed by a path form $d$ to $b$. Since the two sums, the one including all the paths from $c$
to $b$ and the one with all the paths from $d$ to $b$ give the same numerator, the one appearing in the case of
the auxiliary field propagator, the fermionic two-point function is the same one multiplied by $\dirp_{a,c} +
\dirp_{a,d}$, which is $\dirp_{b,a}$, the contraction of the external momentum by Dirac matrices, through
momentum conservation.

Similarly, the Bosonic propagator has either an auxiliary field on the edge $(a,c)$, an auxiliary field on
the edge $(a,d)$ or a fermion loop including both edges. In all three cases, the sum on the configurations
of the other edges will give the common numerator of the graph, multiplied respectively by $p_{a,c}^2$,
$p_{a,d}^2$ and~$-2p_{d,a}\cdot p_{a,c}$. The sum of these three factors is the exterior momentum squared,
ending the general proof of supersymmetry at the integrand level. The situation would be more involved if we
were dealing with the massive case. Indeed, mass terms change the chirality of the fermion, convert scalars to auxiliary
fields, so that new topologies become possible.\looseness=-1

\subsection{Result for the three loop primitive divergence} \label{K33}

The first correction to the Schwinger--Dyson equation for the Wess--Zumino model happens at~3 loop order. The
completion of the relevant graph is the highly symmetric complete bipartite graph $K_{3,3}$, as already pointed
out in~\cite{BeSc12}. Its symmetry
group is indeed the product of two permutation groups $S_3$ acting independently on the two subsets of
vertices.\footnote{As a graph, an additional symmetry exists, exchanging the two subsets of vertices. In the Wess--Zumino model however, the two
subsets are associated to complex conjugated terms in the Lagrangian and are not equivalent.} The numerators we want to compute will share
this high level of symmetry, due to the invariance property proved in the preceding subsection. This will give an important check for our computation.

%% {{{
Let us start by fixing some notations. The $K_{3,3}$ is both highly symmetric and non-planar, so that clear
pictorial depiction is not easy. It is better to say that we have vertices of one kind numbered from 1 to~3,
the ones of the other kind numbered from 4 to~6. Then the edge between vertices $i$ of the first kind and~$j$
of the other can be numbered $3 i + j - 6$.

The Symanzik polynomial associated to this diagram becomes, with these notations
\begin{align*}
 U_K ={}& x_1x_2x_4x_5 +x_1x_2x_4x_6 +x_1x_2x_4x_8 +x_1x_2x_4x_9 +x_1x_2x_5x_6 +x_1x_2x_5x_7
 \\
 &+x_1x_2x_5x_9 +x_1x_2x_6x_7 +x_1x_2x_6x_8 +x_1x_2x_7x_8 +x_1x_2x_7x_9 +x_1x_2x_8x_9
 \\
 &+x_1x_3x_4x_5 +x_1x_3x_4x_6 +x_1x_3x_4x_8 +x_1x_3x_4x_9 +x_1x_3x_5x_6 +x_1x_3x_5x_7
 \\
 &+x_1x_3x_5x_9 +x_1x_3x_6x_7 +x_1x_3x_6x_8 +x_1x_3x_7x_8 +x_1x_3x_7x_9 +x_1x_3x_8x_9
 \\
 &+x_1x_4x_5x_8 +x_1x_4x_5x_9 +x_1x_4x_6x_8 +x_1x_4x_6x_9 +x_1x_5x_6x_8 +x_1x_5x_6x_9
 \\
 &+x_1x_5x_7x_8 +x_1x_5x_7x_9 +x_1x_5x_8x_9 +x_1x_6x_7x_8 +x_1x_6x_7x_9 +x_1x_6x_8x_9 \\
 &+x_2x_3x_4x_5 +x_2x_3x_4x_6 +x_2x_3x_4x_8 +x_2x_3x_4x_9 +x_2x_3x_5x_6 +x_2x_3x_5x_7 \\
 &+x_2x_3x_5x_9 +x_2x_3x_6x_7 +x_2x_3x_6x_8 +x_2x_3x_7x_8 +x_2x_3x_7x_9 +x_2x_3x_8x_9
 \\
 &+x_2x_4x_5x_7 +x_2x_4x_5x_9 +x_2x_4x_6x_7 +x_2x_4x_6x_9 +x_2x_4x_7x_8 +x_2x_4x_7x_9
 \\
 &+x_2x_4x_8x_9 +x_2x_5x_6x_7 +x_2x_5x_6x_9 +x_2x_6x_7x_8 +x_2x_6x_7x_9 +x_2x_6x_8x_9
 \\
 &+x_3x_4x_5x_7 +x_3x_4x_5x_8 +x_3x_4x_6x_7 +x_3x_4x_6x_8 +x_3x_4x_7x_8 +x_3x_4x_7x_9
 \\
 &+x_3x_4x_8x_9 +x_3x_5x_6x_7 +x_3x_5x_6x_8 +x_3x_5x_7x_8 +x_3x_5x_7x_9 +x_3x_5x_8x_9 \\
 &+x_4x_5x_7x_8 +x_4x_5x_7x_9 +x_4x_5x_8x_9 +x_4x_6x_7x_8 +x_4x_6x_7x_9 \\
 &+x_4x_6x_8x_9 +x_5x_6x_7x_8 +x_5x_6x_7x_9 +x_5x_6x_8x_9.
\end{align*}
If we single out the edge~9 between vertices 3 and~6, the numerator comes from the loop $(1,4,5,2)$, with either two of
the edges with an auxiliary field or the entire loop as a fermionic loop. The numerators coming from the auxiliary fields
are easy to obtain as derivatives of the Symanzik polynomial~$U_K$. Using the notation $\partial_i$ for the
derivation with respect to~$x_i$, the numerators corresponding to these cases are simply $\partial_1 U_K \partial_5
U_K + \partial_2 U_K \partial_4 U_K$ for the eight dimensional case, and $\partial_1 \partial_5 U_K +
\partial_2 \partial_4 U_K$ for the six dimensional one. Using the results of Section~\ref{fourdirac}, the fermionic
numerator will need the forest polynomials $U^{1,2,\{4,5\}}$ and~$U^{4,5,\{1,2\}}$. The link between 4 and~5 in the
first one is necessarily over the vertex~3, so that we have to separate the vertices 1, 2 and~3, all linked to vertex~6
through the edges 7, 8 and~9. This allows us to compute the first forest polynomial as well as the second one which is similar
\[
	U^{1,2,\{4,5\}} = x_7 x_8 + x_7 x_9 + x_8 x_9, \qquad
	U^{4,5,\{1,2\}} = x_3 x_6 + x_3 x_9 + x_6 x_9.
\]
The numerator for the term with $U_K^3$ in the denominator can then be computed, remembering that due the chirality
condition, the trace of the identity in spinor space is~2, as
\begin{align*}
	N3 ={}& \partial_1 \partial_5 U_K + \partial_2 \partial_4 U_K - 2\bigl( U^{1,2,\{4,5\}} + U^{4,5,\{1,2\}}\bigr)
	 \\
		={}& x_1x_5+x_1x_6+x_8x_1+x_9x_1+x_4x_2+x_2x_6+x_7x_2+x_2x_9+x_4x_3 \\
		& +x_5x_3+x_7x_3+x_8x_3+x_8x_4+x_9x_4+x_7x_5+x_5x_9+x_7x_6+x_8x_6.
\end{align*}
This expression has the full symmetry of the $K_{3,3}$ graph. Indeed, the 36 different
pairs of edges of this graph fall in only two classes under the action of the symmetry group, the pairs with a~common
vertex and the other ones, each with 18 elements, and $N3$ is the sum of the products of variables associated to the
pairs in the second group. Furthermore, this structure prevents any subdivergence in the corresponding integral. These two
properties of independence on the choice of a special edge and absence of subdivergences must both hold for the complete
integrand, due to the properties established in the previous subsection, but it is remarkable that our choices make them hold for the two terms independently.

As can be expected, the numerator for the $U_K^4$ is more complex, due to its higher degree. Nevertheless, since we
only need one of the forest polynomial $U^{4,5,\{1,2\}}$ and derivatives of the Symanzik polynomial $U_K$, it is not
so hard to compute with an algebraic manipulation program. Indeed the correlation $U^{1,2}$ and $U^{4,5}$, being correlations of neighbouring
edges, can be expressed from derivatives of the polynomial~$U_K$, using that $2 p_1\cdot p_2 = p_1^2 + p_2^2 - p_3^2$. In the Gaussian
integration on the graph, each of the terms give $U^{1,2}$, $\partial_1 U_K$, $\partial_2 U_K$ and $\partial_3 U_K$ and one can deduce the
following relation between these polynomials
\[
	U^{1,2} = \frac12 (\partial_1 U_K + \partial_2 U_K - \partial_3 U_K).
\]
A more graphical proof is possible, grouping the terms
in~$U_K$ with respect to their dependence on $x_1$, $x_2$, $x_3$ as is done in \cite[Lemma~22]{BrSc10}. A similar formula can be given
for~$U^{4.5}$.
\begin{align*}
	N4 ={}& \partial_1 U_K \partial_5 U_K + \partial_2 U_K \partial_4 U_K - 2\bigl(U_K U^{4,5,\{1,2\}}
	\nonumber\\
		& + \frac12 ( \partial_1 U_K + \partial_2 U_K - \partial_3 U_K )
		\frac12 ( \partial_4 U_K + \partial_5 U_K - \partial_6 U_K ) \biggr).
\end{align*}
The resulting expression, of degree 6, has 729 terms and is too large to print here, but can be found in our supplementary
material in file \emph{FZ$33$}. Using the symmetry of the expression, a~description with words is possible, the interested reader can
find it in Appendix~\ref{N4}.

\subsection{Setting for the four loop primitive divergence}\label{section6.3}

At the following perturbative order, there is also a single contributing diagram with a highly symmetric completion, the
cube. In this case, it will be convenient to have a graphical representation of the diagram, even if the plane drawing does
not have the full symmetry of the diagram. The indexes of the edges are chosen in order to make explicit a symmetry of our
computation\looseness=1
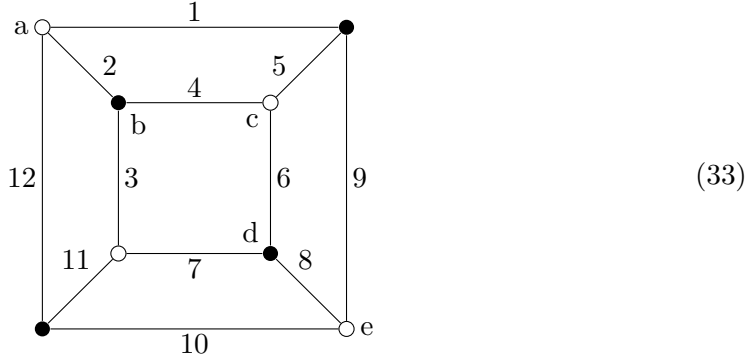
\begin{gather}	\label{cube}
\begin{tikzpicture} [auto]
\node (ehg) at (-2,2) [bla,label=left:a] {} ;\node (ehd) at (2,2) [noi] {};\node (ebd) at (2,-2) [bla,label=right:e] {};
\node (ebg) at (-2,-2) [noi] {};
\node (ihg) at (-1,1) [noi,label=-45:b] {};\node (ihd) at (1,1) [bla,label=-135:c] {};
\node (ibd) at (1,-1) [noi,label=135:d] {};\node (ibg) at (-1,-1) [bla] {};
\draw (ehg) -- node{1} (ehd); \draw (ehg) -- node[near end]{2} (ihg); \draw (ihg) -- node {3} (ibg);
\draw (ihg) -- node{4} (ihd); \draw (ihd) -- node[near start]{5} (ehd); \draw (ihd) -- node {6} (ibd) ;
\draw (ibd) -- node{7} (ibg); \draw (ibd) -- node[near start]{8} (ebd); \draw (ebd) -- node[swap] {9} (ehd) ;
\draw (ebd) -- node{10} (ebg); \draw (ebg) -- node[near end]{11} (ibg); \draw (ebg) -- node {12} (ehg);
\end{tikzpicture}.
\end{gather}
The generic way of determining the numerator would be to choose one particular edge, say~1, as the special one, with all
its neighbours bosonic. The possible decorations of the diagram are therefore a fermionic loop of six edges
(3, 4, 6, 8, 10, 11), two with a fermionic loop of four and one auxiliary field (loop 3, 4, 6, 7 and 10 or loop 7, 8, 10, 11 and 4)
and three with three auxiliary fields (4, 8, 11 and 3, 6, 10 and 4, 7, 10). However, the same numerator can be obtained with
only one fermionic loop of six and two configuration of three auxiliary fields, involving only the even numbered edges,
giving a simpler path to the numerator. This has the added advantage of making a six-fold symmetry explicit.

Let us start with the fermionic loop, $\Tr( \dirp_2\dirp_4\dirp_6\dirp_8\dirp_{10}\dirp_{12}) $.
Then $p_2$ and $p_{12}$ can be expressed through momentum conservation respectively as $p_4 - p_3$ and $p_{10} -
p_{11}$. We therefore obtain four terms, three of which can be simplified from the identity ${\not q \not q} = q^2
\mathbb I$. The trace of the product of two terms just gives the scalar product with a factor 2 due to the chiral trace,
${\Tr ( \not q \not r)} = 2 q\cdot r$, while the identity $p_6 = p_7 + p_8$ can then be used to express the trace
of products of 4 terms with closed loops. We therefore have
\begin{gather*}
 \Tr( \dirp_2\dirp_4\dirp_6\dirp_8\dirp_{10}\dirp_{12}) \\
\qquad = 2 p_4^2 (p_6\cdot p_8) p_{10}^2 - p_4^2 \Tr(\dirp_6\dirp_8\dirp_{10}\dirp_{11})
	- \Tr(\dirp_3\dirp_4\dirp_6\dirp_8) p_{10}^2
+ \Tr( \dirp_3\dirp_4\dirp_6\dirp_8\dirp_{10}\dirp_{11}) \\
 \qquad= 2 p_4^2 p_{10}^2 (p_6\cdot p_8) - 2 p_4^2 p_8^2 (p_{10}\cdot p_{11})
- p_4^2 \Tr(\dirp_7\dirp_8\dirp_{10}\dirp_{11}) \\
\phantom{\qquad =}{} - 2 (p_3\cdot p_4) p_6^2 p_{10}^2 + p_{10}^2 \Tr(\dirp_3\dirp_4\dirp_6\dirp_7)
	+ \Tr( \dirp_3\dirp_4\dirp_6\dirp_8\dirp_{10}\dirp_{11}) .
\end{gather*}
We recognize in the final form the loop of six propagators as well as the two terms with a loop of 4 propagators in the
other formulation of this numerator. The differing sign before one of the loop comes from the reversal of $p_7$ in this
loop in our conventions. Three terms with a scalar product remain. But we know that we must apply the same identity on
$p_2$ and $p_{12}$ in the two terms coming from auxiliary fields, $p_2^2 p_6^2 p_{10}^2$ and $p_4^2 p_8^2
 p_{12}^2$. The expansions of $p_2^2$ and $p_{12}^2$ give three terms for each: the scalar products give terms which
compensate two of these three terms, one of the squared momenta in each case gives one of the auxiliary field terms in the
original formulation and we are left with two terms having the factor $p_4^2 p_{10}^2$. These two terms and the
remaining term with a scalar product in the expansion of the fermionic loop combine to give the last one of the terms with
only auxiliary fields $p_4^2 p_7^2 p_{10}^2$.

\subsection{Evaluation of the numerators}
The Symanzik polynomial $U_C$ for the cube will not be explicitly written, it will simply be available in the
additional files as $Gcu$, since it has 384 terms.

The numerator for the term of dimension~6 is only of degree~2 and is quite easy to compute. The bosonic terms are obtained
from third derivatives of the Symanzik polynomial and only two different kinds of forest polynomials contribute to the
fermionic loop. Indeed, opposing points in the fermionic loop cannot be joined by paths avoiding this loop, so that only
forest polynomials of types $1)$ and~$4)$ are non null. The three polynomials of type~$4)$ are just the product of
the variables associated to two diametrically opposed odd edges, giving $x_1 x_7$, $x_3 x_9$ and~$x_5 x_{11}$. Their
sum will be denoted as $S$. The odd polynomial of type~$1)$ is the symmetric polynomial of degree two
on~$(x_1,x_5,x_9)$ denoted $O$ and the even one is based on the remaining odd numbered variables, denoted $E$
$O = x_1 x_5 + x_1 x_9 + x_5 x_9$, $
	E = x_3 x_7 + x_3 x_{11} + x_7 x_{11}$.
The numerator for the dimension~6 term can then be written as the sum of the bosonic terms, easily expressed by derivation
of the Symanzik polynomial and $-2$ times the listed forest polynomials
\begin{align*}
	N3c ={}& \partial_2 \partial_6 \partial_{10} U_C + \partial_4 \partial_8 \partial_{12} U_C
		- 2 \bigl( S + O + E \bigr) \\
		={}& x_1 x_3 +x_1 x_4 +x_1 x_6 +x_1 x_8 +x_1 x_{10} +x_1 x_{11} +x_2 x_5 +x_2 x_6 +x_2 x_7 +x_2 x_9 \\
		& +x_2 x_{10} +x_2 x_{11} +x_3 x_5 +x_3 x_6 +x_3 x_8 +x_3 x_{10} +x_3 x_{12} +x_4 x_7 +x_4 x_8 \\
		& +x_4 x_9 +x_4 x_{11} +x_4 x_{12} +x_5 x_7 +x_5 x_8 +x_5 x_{10} +x_5 x_{12} +x_6 x_9 +x_6 x_{10} \\
		& +x_6 x_{11} +x_7 x_9 +x_7 x_{10} +x_7 x_{12} +x_8 x_{11} +x_8 x_{12} +x_9 x_{11} +x_9 x_{12}.
\end{align*}
The result is highly symmetric, with 36 different terms characterized by the fact that the two edges associated to the
variables are neither adjacent nor diametrically opposed. It is pitifully the last polynomial that I can explicitly
present in this paper, since the two other ones are far too big.

For the next term, we will need additional forest polynomials. Even for the bosonic terms, some additional complexity
appears since we have three terms from each of the possible field configurations, corresponding to the choice of the
indices in the second derivative. Choosing only one of them, the two other ones are easily deduced by an order three
rotation, easily represented by a shift of 4 in the indices.

For the fermionic part, we also need the terms involving a nearest neighbour correlator.
Again, we only write explicitly the one involving the correlator $U^{10,12}$, knowing there will be two other ones
obtained by rotation. This correlator will be multiplied by the two Dodgson polynomials~${A = U^{2 4,8 6}}$ coming with a
factor 2 to complete the terms of type~$c)$ as well as~${B=U^{2 8,6 4}}$ for the terms of type $b)$. These two terms
will be expressed through a new set of forest polynomials and will use the labels of vertices in~\eqref{cube} to
present them. If we put the two end points $a$ and~$e$ of the path $(2 4 6 8)$ in the same set, we would be back
to the situation for the four-propagator loop and end up with the forest polynomials $U^{\{a,e\},\{b,d\},c}$ for $A$
and~$U^{\{a,c,e\},b,d}$ for $B$, both with a plus sign, but other ones are possible. In the case of Dodgson polynomial
$A$, one can also have crossed associations giving the polynomial $U^{\{a,d\},\{b,e\},c}$ with a~minus sign. However for
the cube, this polynomial is zero, since there is no way to have paths from $a$ to $d$ and $b$ to $e$ which do not
cross. In the case of $B$, there are two additional polynomials, $U^{\{a,c\},\{b,e\},d}=x_7 x_9 x_{12}$
and~$U^{\{a,d\},b,\{c,e\}} = x_1 x_3 x_{10}$.

The numerator for the dimension 8 term will be expressed as the sum of the three images by rotation of the polynomial
$N4p$ minus a contribution written as the product of the Symanzik polynomial $U_C$ by $6 O + 2 S$. We have
\begin{align*}
 N4p ={}& \partial_{10} U_C \partial_2 \partial_6 U_C + \partial_{12} U_C \partial_4\partial_8 U_C -
	2 U^{10,12} (2 A + B) \\
	={}& \partial_{10} U_C \partial_2 \partial_6 U_C + \partial_{12} U_C \partial_4\partial_8 U_C -
		(\partial_{10} U_C + \partial_{12} U_C - \partial_{11} U_C) \cdot \\
	&	( 2 \bigl( x_5 x_{11} (x_1 + x_9 + x_{10} + x_{12}) + x_5 x_{10} x_{12} + x_1 x_9 x_{11} \bigr) \\
	& +	(x_3 x_7 +x_3 x_{11} + x_7 x_{11}) ( x_1 + x_9 + x_{10} + x_{12}) + x_{10} x_{12}(x_3 + x_7) \\
	& + U^{\{a,c\},\{b,e\},d} + U^{\{a,d\},b,\{c,e\}} \bigr).
\end{align*}
Adding its two images by rotation and removing the terms proportional to~$U_C$, we get the numerator~$N4c$.
With 6516 terms with coefficients ranging from 2 to 24, it even defies the kind of description we gave for~$N4$ in
Appendix~\ref{N4}. Indeed, if we could give this kind of description for the 168 terms with coefficient~2, which have
the type~$x^2y^2z^2t$, the 2232 terms with coefficient~3 or the 2976 ones with coefficient~6 are best left to the
supplementary files.

Finally, all elements are available for the description of the degree~12 numerator for the term with $U_C^5$ in the
denominator. It is paradoxically easier to describe than the previous one, since the $B$-type Dodgson polynomials do not
contribute
\begin{align*}
	N5c ={}& \partial_2U_C \partial_6U_C \partial_{10}U_C
	+ \partial_4U_C \partial_8U_C \partial_{12}U_C -2 U^{2,4}U^{6,8}U^{10,12} \\
	& - 2 U_C^2 O - 2 U_C \bigl( U^{10,12} A + \mbox{rotated} \bigr).
\end{align*}
The resulting polynomial has again the full cubic symmetry, but with its $207 359$ terms and coefficients ranging from 1 to
2064, it is beyond simple description. The coefficient 2064 pertains to the single term with the product of all edge
variables, but there are 136 distinct numerical coefficients, 12 of which are common to more than 5232 terms. All these
polynomials are available in the supplementary file \emph{Fcube}.
%% }}}

\section{Conclusion and perspectives}
%% {{{
We have presented a way to obtain simpler parametric representations for graphs with numerator. An important aspect is the
use of the completion of the graph in the case of propagator style graphs, which allows for a higher symmetry of the result
and also limit the highest power of the graph polynomial appearing in the denominator. Knowing the large size of the
obtained numerators, this
claimed simplicity is not so obvious, but we must not forget that, without the reduction to the completed graph, there
would be terms with up to the eighth power of the Symanzik polynomial in the denominator with the concomitant high degree
of the numerator. The simplicity is also in the small number of low degree forest polynomials used as building blocks of
the numerators apart from the derivatives of the Symanzik polynomial.

Our explicit results for the
massless Wess--Zumino model should be essential stepping stones for the explicit study of the corrections to the
Schwinger--Dyson equation of this model. In~\cite{BeSc08}, only the lowest term in the equation was included, but already
in~\cite{BeSc12} the effect of these three and four loop corrections to the Schwinger--Dyson equations on the asymptotic
properties of the perturbative series was studied. However, it was only a negative result that the leading asymptotic
behaviour should not be changed. With the full parametric representation of these diagrams, explicit corrections to this
asymptotic behaviour could be computed as well as the subleading contributions to the $\beta$-function itself.
One may wonder if the high number of terms in the obtained numerators would preclude the practical
use of such expressions. Combined with the fifth power of the Symanzik polynomial in the denominator, this could be
stretching the capacity of the presently available programs for the analytic integration of such expressions. However, the
high symmetry and the avoidance of any spurious subdivergence should make this difficulty less serious. Moreover, some of
the important use cases of these expressions as the determination of the residues of the Mellin transform of the
corresponding graphs will only need the terms with some given power of a variable, selecting a reduced number of
terms. The combination of exact supersymmetry, possible since we remain in four dimensions, and explicit cancellation of
all subdivergences gives tremendous advantages over the use of dimensional regularisation.

The results obtained for the fermionic loops can also be used for five or six loops primitive diagrams, but only for their lowest
order contribution: we hope to show in a future publication how to compute the full five or even six loops terms in the
$\beta$-function of the Wess--Zumino model. Having a full description of the five loop primitives would however be more
demanding: in this case, we would need either the case of two independent loops with four fermionic propagators or the one
of the eight fermionic propagator loops. Both cases introduce additional hurdles for their evaluation, in particular for
the case of two independent loops for which the chirality constraint is more than just halving the dimension of the spinor
space. And there is the additional question of dealing with the possibly hundreds of millions of terms in the obtained
numerators.

Applications of this formalism to other quantum field theories would also be interesting. In the case of Yang--Mills
theories, a path towards parametric representations of arbitrary diagrams (or rather, of the collection of diagrams sharing
a common color structure) has been made through the introduction of corolla polynomials~\cite{KrSaSu12,KrYe12}, but
additional variables remain. An explicit evaluation of the effect of the differential operators present in this formalism
to obtain expressions involving only the parameters associated to edges would be interesting, especially if the complete
result can be shown to have a high degree of symmetry. Fermion loops in gauge theories would also add to the complexity,
due to the gamma matrix appearing in the vertex. Finally, we limited ourselves to the massless case, since it allows for
the reduction to the completed graph with the simple proof of Section~\ref{reducgene}. Nevertheless, similar identities
could be used: terms with the highest possible number of $V^i$ terms are still the ones with the highest power of the
first Symanzik polynomial in the denominator and any way of reducing this power would be welcomed.

In the Wess--Zumino model our examples are based on, combining the numerators for the differing assignations of fields to
the edges of the graph made perfect sense, since it allowed to avoid all spurious divergences and the computations are
rather straightforward with respect to the reductions based on IBP identities. Nevertheless, reducing the
total number of different integrals would be a bonus. It would be interesting to find ways of combining the
explicit description of numerators in the parametric representation presented in this work with reduction methods to try to
get the best of both worlds.

\appendix
\section{Dodgson identities}\label{polyn}

In order to obtain exact results, we need the Dogdson identities in a form which allows to track signs.
They are indeed essential to obtain exact results. In other contexts, the signs can be irrelevant
since the Dodgson polynomials will appear squared or as part of expressions the sign of which is not important,
but it is not the case here. Nevertheless, we want to keep things simple and with
signs which depend on a minimum number of choices in the presentation of the diagram. In particular, the usual
presentation as the determinant of a submatrix with rows and columns removed keeps a dependence on the order
of the rows. The convention we proposed in Section~\ref{presentation} for $U^{I ,J}$ is the determinant of the matrix
$M_G$ with the columns indexed by the set $I$ replaced with ones with only a 1 in the position indexed by the corresponding
element of~$J$. Independence on the order of the edges or the vertices is clear in this case and expansion according to
the columns indexed by~$I$ trivially gives back the usual definition up to a sign, while multilinearity can be used to
show that the same result can be obtained by changing instead the rows indexed by~$J$. Finally, in the case where $I$
and~$J$ have only one element, $U^{i,j}$ only depends on the orientations of $i$ and~$j$.

The sign introduced can be deduced as follows: the elements of $I$ have first to be brought to the beginning of the matrix,
requiring $i_1 - 1$ transpositions to send the first element of $I$ to~1,~${i_2 -2}$ for the following up to $i_k -k $
for the last one if we suppose that the elements of~$I$ are numbered according to the ordering of edges and not the order in~$I$.
We have similar number of transpositions for bringing the elements of $J$ to the front, and the total sign only depends on the parity
of the sum $\sum_{i\in I} i+ \sum_{j\in J} j$. The upper left block of the resulting matrix is a~permutation matrix, which gives a sign
corresponding to the product of the signs associated to the permutations bringing $I$ and~$J$ in the natural order. If one recalls that
the convention in this work is to put a minus sign before one of the incidence block rather than to use the symmetric form used by Schnetz,
one recognizes that there are all the elements of the sign proposed in \cite[equation~(8)]{SC2021}.

Our aim is to express some determinants involving correlators~$U^{i,j}$. Examining the proof of Dodgson
identities, such
as the one found in~\cite{Br2009}, shows that our definition of~$U^{I ,J}$ naturally appears in them. Indeed, $U^{i,j}$
is the $j,i$ component of the adjugate matrix of $M_G$, the matrix which, multiplied by $M_G$ gives $\det(M_G)$
times the identity. We want to express the determinant of the matrix $M$ with elements $U^{i,j}$ such that $i$ is in $I$ and $j$ is in
$J$. We form a matrix of the dimension of~$M_G$ by replacing in the identity matrix the columns with index in~$J$ by the
ones indexed by~$I$ of the adjugate matrix. An expansion along the columns coming from the identity matrix shows
that this new matrix has the same determinant as~$M$.

Multiplying by the left by~$M_G$, one obtains a matrix which has the columns of~$M_G$ except for the columns indexed
by~$J$ which have $\det(M_G)$ in the position indexed by~$I$. We recognize the matrix used to define $U^{J,I}$,
apart from the factors $\det(M_G)$, so that we obtain the required Dodgson identities without any sign ambiguity by
taking the determinants
\[
	\det(M_G) \det\bigl( (U^{i,j})_{i\in I,j\in J} \bigr) = \det(M_G)^{\# J} U^{I ,J}.
\]

\section[Description of N4]{Description of $\boldsymbol{N4}$} \label{N4}

We use here the high level of symmetry of $N4$ to give it a not so large verbal description. This description also
shows that none of the terms can give a contribution with subdivergences. The possible terms are constrained by the facts
that the degree in each variable is at most 2
and the combined powers of the variables linked
to the three edges meeting at a given vertex cannot be greater than~3. This in particular precludes that the edges
associated to two squared variables may meet in a common vertex.

We first have 6 terms of the type $x^2 y^2 z^2$, with coefficient 1: the three corresponding edges do not have any vertex in
common. There are $18 \times 15$ terms of type $x^2 y^2 z t$, still with coefficient~1: there are 18 choices of the
edges associated to $x$ and~$y$ which have no common vertex, as in the polynomial~$N3$. Among the 21 pairs of edges
possible among the 7 remaining ones, 6 are excluded: the four pairs which have one of the ends of $x$ or $y$ as a
common vertex and the two pairs where $x$ and~$z$ have a common vertex, $y$ and~$t$ also, but there are no other
common vertices between the four edges.

There are then $9\times 41$ terms with exactly one square. They have coefficient~2.
The 9 factor comes from the 9 ways of choosing the edge which corresponds to the squared variable. It remains to choose
4~edges among the remaining eight ones. They are divided between the four edges having a vertex in common with the one
associated with the squared variables and the remaining four. The basic rule enunciated in the previous paragraph then
limits to two the number of edges in the first group. There is a first possibility with the four edges in the second group,
4 times 4 possibilities where we choose one edge in the first group and exclude one in the second and finally 4 times 6
where we choose one edge from each of the end of the squared edge and the last two among the four of the second group for
the announced total of 41 possibilities.

Finally, the 84 possible terms with six different variables are all present, but
they come with coefficients 8, 6 or~4. It is easier to describe these terms by the variables which are {\em not} included.
The 6 terms where the excluded variables correspond to edges without common vertices have coefficient~8, the 36 ones where
the excluded edges form an unbranched path have coefficient 4 and the remaining 42 ones have coefficient~6.

\subsection*{Acknowledgements}

I would like to thank the anonymous referees for their critics and suggestions. They addressed serious shortcomings
in previous versions of this work to make it more readable, both in terms of the precise results established and the
way they can be derived. Any remaining difficulties are nevertheless of my entire responsibility.

\pdfbookmark[1]{References}{ref}
\LastPageEnding

\end{document}